\documentclass[12pt]{article}
\usepackage[dvips]{graphics}
\usepackage{amsfonts}
\usepackage{amssymb}

\def\be{\begin{eqnarray}}
\def\ee{\end{eqnarray}}
\def\bee{\begin{eqnarray*}}
\def\eee{\end{eqnarray*}}

\def\ra{\rangle}
\def\la{\langle}

        \def\iff{\Leftrightarrow}
   \def\imp{\Rightarrow}
        \def\tr{\hbox{Tr}}
        
      \def\Hil{{\cal H}}
\def\co{{\cal O}}
\def\bu{{\bf u}}
\def\bv{{\bf v}}
\def\bV{{\bf V}}
\def\bw{ {\bf w}}

\def\ot{\otimes}
\def\shan{{\rm Shan}}
\def\holv{{\rm Holv}}
\def\pure{{\rm pure}}
\def\DM{\rm DenMat}

\def\bra{\langle}
\def\ket{\rangle}
\def\kb{ \ket \bra }

\def\rt2{ \frac{1}{\sqrt{2}} }

\def\raw{\rightarrow}

\def\wh{\widehat}
\def\dg{\dagger}
\def\dtsig{{\mathbf \cdot \sigma}}
\def\rmD{{\rm D}}

\def\half{{\textstyle \frac{1}{2}}}
\def\frth{{\textstyle \frac{1}{4}}}
\def\thrd{{\textstyle \frac{1}{3}}}

\def\nl{\newline}

\def\rmT{{\rm T}}
\def\BbbT{{\Bbb T}}

\newtheorem{thm}{Theorem}
\newtheorem{cor}[thm]{Corollary}
\newtheorem{lemma}[thm]{Lemma}
\newtheorem{conj}[thm]{Conjecture}

\title{
Minimal Entropy of States  \\ Emerging from Noisy Quantum Channels}

\author{Christopher King \thanks{Partially supported by National Science
        Foundation Grant DMS-97-05779 } \\ Department of
        Mathematics \\ Northeastern University \\
 Boston, MA 02115  \\ {\normalsize king@neu.edu} \and
Mary Beth Ruskai \thanks{Partially supported by National Science
        Foundation Grant DMS-97-06981 and Army Research Office Grant
   DAAG55-98-1-0374} \\ Department of
        Mathematics \\ University of Massachusetts  Lowell \\ Lowell,
        MA  01854 USA \\ {\normalsize bruskai@cs.uml.edu} }

\begin{document}

\maketitle

\begin{abstract}
In this paper, we consider the minimal entropy of qubit states
transmitted through two uses of a noisy quantum channel, which is
modeled by the action of a completely positive trace-preserving
(or stochastic) map. We provide strong support for the conjecture
that this minimal entropy is additive, namely that the minimum
entropy can be achieved when product states are transmitted.
Explicitly, we prove that for a tensor product of two unital stochastic
maps on qubit states, using an entanglement that involves only states
which emerge with minimal entropy cannot decrease the entropy below the
minimum achievable using product states. We give a separate argument,
based on the geometry of the image of the set of density matrices
under stochastic maps, which suggests that the minimal entropy
conjecture holds for non-unital as well as for unital maps.  We
also show that the maximal norm of the output states is
multiplicative for most product maps on $n$-qubit states, including all
those for which at least one map is unital.

For the class of {\it unital} channels on ${\bf C}^2$, we show
that additivity of minimal entropy implies that the Holevo
capacity of the channel is {\it additive} over two inputs,
achievable with orthogonal states, and equal to the Shannon
capacity.   This implies that superadditivity of the capacity is
possible only for non-unital channels.
\end{abstract}

\bigskip
\noindent{\it Key words:} Entangled state; minimal entropy;
qubit; stochastic map.

\tableofcontents

\pagebreak

\section{Introduction}

\subsection{Entropy and unital stochastic maps}

When a pure state, represented by a density matrix $\rho$, is
transmitted along a noisy channel, it is mapped into a mixed
state $\Phi(\rho)$.  The entropy of the initial pure state
is necessarily zero, i.e.,
$S(\rho) \equiv - \tr \, \rho \, \log \rho = 0$ since
${\rho}^2 = \rho$ and so the only
eigenvalues of $\rho$ are $0$ and $1$.  However, the
entropy $S[\Phi(\rho)]$ of the mixed state which emerges need
not be zero.   One seeks states $\rho$ which minimize the
effect of the noise in the sense of minimizing the entropy
$S[\Phi(\rho)]$ of the state that emerges from the channel.
There are a number of reasons for studying such states,
most notably the  connection between minimizing entropy
and maximizing channel capacity, which will be
discussed in Section \ref{subsect:prelim.cap}.  However, in this paper
we focus attention on the entropy.

The noise, which results from interactions between the states
in a Hilbert space $\Hil$
and the environment, is represented by the action of a
completely positive, trace-preserving map $\Phi$ on the
trace class operators in ${\cal B}(\Hil)$.
We use the term {\em stochastic} to describe such maps.
(Following a similar use by Alberti and Uhlmann \cite{AH},
this terminology was used by Petz,
\cite{Pz} and reflects the fact that $\Phi$ is the non-commutative
analogue of the action of a column stochastic matrix on a probability
vector.)  We restrict attention to two-level quantum systems in which
case $\Hil = {\bf C}^2$ or tensor products of ${\bf C}^2$.  A stochastic
map $\Phi$ acting on states in ${\bf C}^2$, can be naturally extended to
tensor products, e.g., $\Phi \ot \Phi$ acting on states
on ${\bf C}^2 \ot {\bf C}^2 $ etc.,  and leads to questions about
the additivity of the minimal entropy and capacity of product
channels.

In particular, Shor has conjectured that the minimal
entropy is additive.  We were led independently to this conjecture
because it would imply additivity of channel capacity for {\em unital}
stochastic maps, i.e., maps which take the identity operator
to itself so that $\Phi(I) = I$.  Although we have a convincing argument
that our results for entropy, unlike those for channel capacity,
extend to non-unital maps, we focus most of our attention on
unital maps.  In the last section we briefly consider non-unital maps.

\bigskip

Recall that every completely positive map $\Phi$ can be
represented (non-uniquely) in the Kraus form
\be \label{eq:kraus}
\Phi(\rho) = \sum_k A_{k}^{\dagger} \rho A_{k}.
\ee
Every map $\Phi$ representing noisy evolution in a quantum channel
must preserve the trace of $\rho$, since $\Phi(\rho)$ is also
a state. In terms of the Kraus operators,
the condition that $\Phi$ be stochastic, that is
completely positive and trace preserving, is
\be  \label{eq:cond.tracepres}
\tr \Phi(\rho) = \tr \rho \quad \forall \rho & \iff & \sum_{k=1}^{n} A_{k}
A_{k}^{\dagger} = I .
\ee
The map $\Phi$ is {\it unital} if $\Phi(I) = I$, that is
if $\Phi$ maps the identity operator to itself.
In terms of the Kraus operators,
the  condition  for $\Phi$ to be completely positive and unital is
\be  \label{eq:cond.unital}
\Phi(I) = I ~~ & \iff &
\sum_{k=1}^{n} A_{k}^{\dagger} A_{k} = I  .
\ee
A sufficient condition that a stochastic map be unital is that the
Kraus operators are self-adjoint, i.e., $A_k = A_k^{\dagger} ~ \forall k$.
This condition is not necessary; for example, a
double-stochastic matrix which is not symmetric corresponds to a
unital stochastic map which is not self-adjoint.
Henceforth in this paper, ``unital map'' will mean ``unital stochastic map''
unless otherwise stated.

It is worth noting that the Kraus operators are self-adjoint if
and only if $\Phi$ is self-adjoint with respect to the
Hilbert-Schmidt inner product
$\bra P, Q \ket = \tr P^{\dagger} Q$.  The
dual or adjoint map $\hat{\Phi}$ is then defined by the
condition $\tr [\hat{\Phi}(P)]^{\dagger} Q = \tr  P^{\dagger} \Phi(Q)$.
It is easy to see that if $\Phi$ is given by (\ref{eq:kraus}) then
$\hat{\Phi}(\rho) = \sum_k A_{k}\rho A_{k}^{\dagger} $.
In addition, since the dual of any trace-preserving map
satisfies $\hat{\Phi}(I) = I$, any stochastic map (considered as
a linear map on the space of Hilbert-Schmidt operators) has
eigenvalue $1$, and hence a fixed point $P$ such that $\Phi(P) = P$.

For unital maps, the identity is an eigenvector whose orthogonal
complement is the set of operators with trace zero.  Hence, a
unital map also defines a linear map on the traceless part
of a density matrix.  By contrast, a non-unital map is only affine
on the set of traceless matrices.  This distinction is easily
seen for the ${\bf C}^2$ case when the Bloch sphere representation
is used as described in section \ref{sect:notation}.
Appendix C contains a list of examples of unital and non-unital maps.


\bigskip

Recall that the entropy of a density matrix can be written in
terms of its eigenvalues $\lambda_k$, namely
$S(\rho) = -\sum_k \lambda_k \log \lambda_k$.  The minimal
entropy $S(\rho) = 0$ occurs if and only if one eigenvalue of $\rho$ is
$1$ and all others $0$; the maximal entropy (in $d$ dimensions)
of $S(\rho) = \log d$ occurs if and only if all eigenvalues
are $1/d$ so that $\rho = \frac{1}{d} I$.  Thus, if
$S(\rho) \approx 0$, one must have one eigenvalue close to $1$
and the others near $0$.  Hence, states with small entropy
are those for which $\| \rho \| \approx 1$.  Thus, in seeking
pure states $\rho$ which have minimal entropy $S[\Phi(\rho)]$
after emerging from a noisy channel, we are led to seek states
for which $\| \Phi(\rho) \|$ is maximal.  In section \ref{sect:norm.bnd}
we give a precise definition of the maximal norm and show that
it is multiplicative when (at least) one channel is a unital map on
${\bf C}^2$.

Our results suggest that multiplicativity of the maximal norm may
hold for general channels; in fact, we can extend our result to some
non-unital channels (see Remarks at the end of Section \ref{sect:norm.bnd}).
In any case, our results provide
strong support for the additivity of minimal entropy for
unital channels.

Another property of unital channels is that the entropy
of a state is non-decreasing under the action of a unital
stochastic map.
This follows easily from the fact that the relative entropy
\be
   H(P,Q) =  \tr P [\log P - \log Q]
\ee
decreases under stochastic maps, i.e.,
\be \label{eq:relent.dec}
 H[\Phi(P),\Phi(Q)] \leq  H(P,Q)
\ee
Since $S(\rho) = -H(\rho, \frac{1}{d} I) + \log d$
if $\Phi$ is unital, it follows from (\ref{eq:relent.dec})
that $S[\Phi(\rho)] \geq S(\rho)$.
For a non-unital map, the entropy of a pure state cannot decrease;
however, one can have mixed states for which the entropy actually
decreases.

\subsection{Channel Capacity}

We now discuss the   information
capacity of a noisy quantum channel \cite{Ho1}, \cite{Ho2} used for
what is sometimes called ``classical'' communication, i.e.,
communications in which signals are sent using quantum particles
but without additional or prior entanglement between sender
and receiver.
In the simplest case where no entanglement
is used in either the transmission or the measurement,
each letter $i$ from the source alphabet is
represented by a pure state  which we represent by its
density matrix ${\rho}_i$  on a quantum Hilbert space.
During transmission
the channel transforms this state into ${\tilde \rho}_i \equiv
\Phi({\rho}_i)$, where $\Phi$  implements the noisy interaction
between states and the environment. The map $\Phi$
is a completely positive, trace-preserving map on the set of states.
The resulting state ${\tilde \rho}_i$ is measured, and the outcome
determines a letter from the output alphabet. In the general case this
measurement is effected
by a positive operator-valued measurement (POVM) -- namely, there is a positive
operator $E_j$ assigned to each letter $j$ of the output alphabet, which
together
satisfy the constraint $\sum_{j} E_j = I$. When the measurement is
performed on a state
$\rho$, the result will be $j$ with
probability ${\rm Tr}\big(\rho E_j\big)$.

Several definitions of channel capacity have been proposed,
corresponding to  whether or not entangled states are used for
transmission, and whether or not entangled measurements are made by the
receiver.  Bennett and Shor \cite{BS} identify four possibilities,
which we denote $C_{PP}, C_{PE}, C_{EP}$ and $C_{EE}$ where
the subscripts $P$ and $E$ refer to product and entangled
processes respectively.

\bigskip

In the process described above, with no entanglement at either end,
the channel is equivalent to a classical
noisy channel with transition probabilities
$\{p_{ij} =  \tr \big({\tilde \rho}_i E_j\big)\}$.
Therefore its
maximum rate of information transmission is given by
\be\label{Shannon}
C_{PP}(\Phi) = C_{\rm Shan}(\Phi) = \sup_{\pi, \rho, E}
\sum_{i} \sum_{j}  {\pi}_i p_{ij} \log \bigg(
{p_{ij} \over \sum_{k} {\pi}_k p_{kj}} \bigg),
\ee
where we take the $\sup$ over all probability distributions $\{{\pi}_i\}$
for the
input alphabet, as well as all choices of input states and measurements.
We call this the {\it Shannon capacity} of the channel, since it bears closest
resemblance to the classical case.

It is reasonable to expect that by transmitting entangled states and by
using entangled
measurements it may be possible to exceed the Shannon capacity for a noisy
channel. The Holevo-Schumacher-Westmoreland Theorem \cite{Ho1}, \cite{SW}
provides a closed form expression for the capacity in the case
where product states are transmitted at the input
and entangled measurements of arbitrary length are allowed at the output:
\be\label{HSW}
C_{PE} = C_{\holv}(\Phi) .
\ee
Here $C_{\holv}(\Phi)$ is the {\it Holevo capacity} of the channel:
\be\label{Holevo}
C_{\holv}(\Phi) = \sup_{\pi, \rho}
\bigg(S({\tilde \rho}) - \sum_{i} {\pi}_i S({\tilde \rho}_i)\bigg),
\ee
where $\rho = \sum_i {\pi}_i {\rho}_i$ and
${\tilde \rho} = \Phi(\rho)$. The well-known Holevo bound states that
\be
 C_{\shan}(\Phi) \leq C_{\holv}(\Phi).
\ee
Holevo \cite{Ho2, Ho3} provided examples of channels in which this
inequality is strict, i.e.,  $C_{\shan}(\Phi) < C_{\holv}(\Phi)$.
Furthermore, it has  been shown \cite {FC,OPW}  that a
necessary and sufficient condition for strict inequality is
that the output states $\{{\tilde \rho}_i\}$ do not commute.

\bigskip

One important open question is whether or not the
Holevo capacity can be exceeded when
entangled states are used at the input, that is whether
$C_{EE}$ exceeds $C_{PE}$. This would be equivalent to the
superadditivity of the Holevo capacity.
In this paper we address this question in the case of messages
which are entangled over {\it two inputs} only. This
is equivalent to the question whether
$C_{\rm Hol}(\Phi \otimes \Phi)$ exceeds $2 C_{\rm Hol}(\Phi)$.
Holevo \cite{Ho2} has shown that $C_{\shan}(\Phi \otimes \Phi) > 2
C_{\shan}(\Phi)$ for the quantum binary channel, but to the
best of the authors' knowledge there is no known example
of a superadditive channel  for the Holevo capacity.
Bruss et al \cite{BFMP} showed that
$C_{\rm Hol}(\Phi \otimes \Phi) = 2 C_{\rm Hol}(\Phi)$ for the
depolarising channel, which is an example of a unital channel.
As we will show, our results strongly suggest that {\it if} the
Holevo capacity is superadditive then the channel must be non-unital.

\subsection{Summary of Results}\label{sect:summary}

We prove several
theorems about the minimal entropy and the maximal norm for
states of the form ${(\Phi \otimes \Omega) ({\rho}_{12})}$, where
 $\Phi$ and $\Omega$ are unital stochastic maps on ${\bf C}^{2\times 2}$
and
${\rho}_{12}$ is an entangled state. In addition, we explain how
these results provide evidence for
the conjecture that {\it minimal entropy is additive} for all stochastic
maps on ${\bf C}^{2\times 2}$.   We also show that this conjecture has
important implications for the capacity of unital quantum channels. In
particular, we show that the conjecture implies that if $\Phi$ is
unital, then the Holevo capacity is additive over two inputs, that is
${C_{\holv}(\Phi \otimes \Phi) = 2 C_{\holv}(\Phi)}$.

\medskip

Our first theorem concerns the maximal value of $\|\Phi(\rho)\|$
as $\rho$ varies over states on ${\bf C}^{2}$.  We will
consider the general possibility of two
stochastic maps $\Phi$ and $\Omega$ on ${\bf C}^{2\times 2}$
and denote their maximal values by $M_{\Phi}$ and $M_{\Omega}$
respectively.  In Theorem \ref{thm:normbnd} we
 prove, under mild conditions on one of these maps, that the maximal value
of
$\|(\Phi \otimes\Omega)({\rho}_{12})\|$ is $M_{\Phi} M_{\Omega}$, as
${\rho}_{12}$ varies over states on ${\bf C}^{2\times 2}$.
That is, the norm
of $(\Phi \otimes \Omega)({\rho}_{12})$ achieves its maximal value on product
states ${\rho}_{12} = {\rho}_{1}
\otimes {\rho}_{2}$, rather than on entangled states.

In Theorem \ref{thm:mixing}, we prove a similar, though slightly weaker,
result for the minimal entropy of $(\Phi \otimes \Omega)({\rho}_{12})$.
Namely, we restrict ${\rho}_{12}$ to the family of entangled states whose
reduced density matrices ${\rho}_{1}$ and ${\rho}_{2}$ are such that
$\Phi({\rho}_{1})$ and $\Omega({\rho}_{2})$ have minimal entropy.
Then we prove that the minimal entropy of $(\Phi \otimes
\Omega)({\rho}_{12})$, as
${\rho}_{12}$ varies over this family, is
 the sum of the minimal entropies
of $\Phi({\rho}_{1})$ and $\Omega({\rho}_{2})$. That is, the entropy also
achieves its minimal value on product states.

It seems extremely unlikely that we could further decrease the entropy of
$(\Phi \otimes \Omega)({\rho}_{12})$
by using an entangled state whose reduced density matrices do not have
minimal entropy themselves.
Therefore we believe that the conclusion of Theorem \ref{thm:mixing}
also holds as ${\rho}_{12}$ varies over all entangled states on
${\bf C}^{2\times 2}$. In fact, similar
arguments and numerical evidence support an even stronger conclusion,
namely that
minimal entropy is additive for {\it all} stochastic maps on
${\bf C}^{2\times 2}$. This is the content of the conjecture below.

\begin{conj} \label{conj:minent}
If $\Phi$ and $\Omega$ are stochastic maps on ${\bf C}^{2\times 2}$, then
\be
  \inf_{\rho~ \pure} S\Big[(\Phi \otimes \Omega)(\rho)\Big] =
    \inf_{\rho~ \pure} S\Big[\Phi(\rho)\Big] +
\inf_{\rho~ \pure} S\Big[\Omega(\rho)\Big].
\ee
\end{conj}
We will discuss the evidence for this conjecture in detail in
Section \ref{sect:entanal} and Section \ref{subsect:nonu.min.ent}.
We note that
P. Shor \cite{Shor} earlier made a similar conjecture, and
together with J. Smolin
obtained numerical evidence which supports it.

\bigskip
The unital case is particularly important because it yields the
following results as immediate corollaries.

\begin{cor}
If $\Phi$ is unital, then the Holevo capacity is
additive, i.e.,\linebreak  ${C_{\holv}(\Phi \otimes \Phi) = 2
C_{\holv}(\Phi)}$.
\end{cor}
\begin{cor} If $\Phi$ is unital, then the Holevo capacity
 can be achieved with orthogonal states.
\end{cor}
\begin{cor}
If $\Phi$ is unital, then $C_{\holv}(\Phi) = C_{\shan}(\Phi)$ and
$C_{\holv}(\Phi \otimes \Phi) = C_{\shan}(\Phi \otimes \Phi)$.
\end{cor}
\begin{cor} If $\Phi$ is unital, then the Shannon capacity is also
additive, i.e., \linebreak ${C_{\shan}(\Phi \otimes \Phi) = 2
C_{\shan}(\Phi)}$.
\end{cor}

In Section \ref{subsect:prelim.cap} we will explain in detail how these
Corollaries follow if Conjecture \ref{conj:minent} holds for unital maps.

\bigskip

This paper is organized as follows.  In Section \ref{sect:prelim} we
introduce the notation we will use for the Stokes parametrization
for representing both
states and maps in a basis consisting of the Identity and
Pauli matrices, and show how the various conditions of
unital, trace-preserving, and complete positivity can be
expressed in this representation.  With this background,
we conclude Section \ref{sect:prelim} by presenting
 the arguments leading
to the Corollaries above.

In Section \ref{sect:norm.bnd} we prove the multiplicativity of
the maximum value of $\| \Phi(\rho) \|$.
Section \ref{sect:entanal} contains the heart of the paper in which we
give the details of the proof of our theorem about additivity
of minimal entropy.
In Section \ref{sect:non.unit} we discuss some of the features of
non-unital maps using a special subclass and then present the evidence for
additivity of minimal entropy in general.
Finally, in Section  \ref{sect:conc} we summarize our results and discuss
their implications for channel design.

We also include three Appendices.  Appendix A gives some
important background on singular value decompositions and
the details needed for the diagonal representation
introduced in Section \ref{sect:notation}.  Appendix B gives the details
needed to verify the complete positivity conditions
of Section \ref{sect:singv.cond}.  In Appendix C we provide a number of
examples of different types of channels and show how some familiar
examples appear in the representation and notation we use.

\section{Preliminaries}\label{sect:prelim}


\subsection{Stokes parametrization and Bloch sphere} \label{sect:notation}

Recall that the identity and Pauli matrices form a basis
for ${\bf C}^{2 \times 2}$ so that any $2 \times 2$ matrix $C$
can be written as $w_0 I + \bw \dtsig $ where
${\bf \sigma}$ denotes the vector of Pauli matrices
and ${\bf w} \in {\bf C}^3.$ Then for $C$ = $w_0 I + \bw \dtsig $
\begin{itemize}
\item[a)] $C$ is self-adjoint $\iff~~(w_0,\bw)$ is real, i.e.,
$w_0 \in {\bf R}$ and $\bw \in {\bf R}^3$;
\item[b)] $\tr C = 1 \iff  w_0 = \half$; and
\item[c)] $ C > 0 \iff  |\bw| \leq  w_0.$
\end{itemize}
Thus, $\{ I, \sigma \}$ also form a basis for the {\em real}
vector space of self-adjoint matrices in ${\bf C}^{2 \times 2}$
and every density matrix can  be written in this basis
as $\rho = {\half} [I + \bw \cdot {\bf \sigma}]$ with
$\bw \in {\bf R}^3$ and $|\bw| \leq 1$.  Furthermore
\begin{itemize}
\item[d)] $\rho$ is a one-dimensional projection (or pure state)
$\iff~~ |\bw| = 1.$
\end{itemize}

Every linear map
 $\Phi : {\bf C}^{2 \times 2} \rightarrow {\bf C}^{2 \times 2}$
can be represented in this basis by a unique $4 \times 4$
matrix ${\BbbT}$, and $\Phi$ is trace-preserving if
and only if the first row satisfies $t_{1k} = \delta_{1k}$, i.e.,
${\BbbT} =  \left( \begin{array} {cc}
     1 & {\bf 0} \\  {\bf t} & {\rmT} \end{array} \right)$
where $\rmT$ is a $3 \times 3$ matrix (and ${\bf 0}$ and ${\bf t}$
are row and column vectors respectively) so that
\begin{eqnarray} \label{eq:Trep}
  \Phi(w_0I + \bw \cdot \sigma) =
   w_0I + ({\bf t} + {\rmT} \bw) \cdot \sigma .
\end{eqnarray}
The matrix ${\BbbT}$ is self-adjoint if and only if $\Phi$ is
self-adjoint as an operator on ${\bf C}^{2 \times 2}$
with respect to the Hilbert-Schmidt inner product.  We are
interested in those $\Phi$ which map a subspace of
self-adjoint matrices into itself, which holds if and
only if $\BbbT$ is real.

The map $\Phi$ is unital if and only if ${\bf t} = {\bf 0}$. Thus,
any unital stochastic map $\Phi$ acting on density matrices on ${\bf C}^2$
can be written in the form
\be \label{eq:T3rep}
\Phi\big( \half [I + \bw \dtsig] \big) =
\half [ I +   (\rmT \bw) \dtsig ],
\ee
where $\rmT$ is a real $3 \times 3$ matrix. Using the singular value
decomposition (see Appendix A), we can write
\be\label{SVD}
\rmT = R S
\ee
where $R$ is a rotation and $S$ is self-adjoint.
Define the map ${\Phi}_{S}$ by
\be
{\Phi}_{S}\big( \half [I + \bw \dtsig] \big) =
\half [ I +   (S \bw) \dtsig ]
\ee
As explained in Appendix A, the rotation $R$ defines a unitary
operator $U$ such that for any state $\rho$,
\be\label{eq:def.U}
\Phi(\rho) = U \Big[{\Phi}_{S} \big( \rho \big)
\Big]U^{\dagger}
\ee
In this paper we are interested only in the critical
values of certain functions of the spectrum of $\Phi(\rho)$,
as $\rho$ varies over the space of states -- the maximum value of
the norm, the minimum value of the entropy. Since a unitary
transformation leaves the spectrum unchanged, these are the same for
$\Phi$ and ${\Phi}_{S}$. Also, since $S$ is self-adjoint it can be
diagonalized by a change of basis.
Hence without loss of generality we need henceforth
consider only unital stochastic maps whose matrix $\rmT$ defined in
(\ref{eq:Trep}) is diagonal, with eigenvalues
$({\lambda}_1, {\lambda}_2, {\lambda}_3)$. As a shorthand, we will denote
this diagonal map by $\Phi[{\lambda}_1, {\lambda}_2, {\lambda}_3]$.
The image of the set of pure state density matrices
$\rho = \half[I + \bw \dtsig ]$ (with $|\bw| = 1$) under the action of
$\Phi[{\lambda}_1, {\lambda}_2, {\lambda}_3]$ is the ellipsoid
\be \label{eq:ellip.unit}
 \left( \frac{ x_1 }{\lambda_1} \right)^2 +
   \left( \frac{ x_2 }{\lambda_2} \right)^2 +
   \left( \frac{ x_3 }{\lambda_3} \right)^2 = 1,
\ee
and the image under the action of $\Phi$ is obtained by a
further rotation of the ellipsoid, corresponding to
the operator $U$ in (\ref{eq:def.U}).

Similar reasoning applies when $\Phi$ is non-unital. Using
(\ref{eq:T.SVD.nonpos}) and (\ref{Phi.selfadj}) from Appendix A, the map
$\Phi$ can be written in the form
$\Phi(\rho) = U \Phi_D (V \rho V^{\dg}) U^{\dg}$ where $U,V$
are unitary, $D$ is diagonal and
$\Phi_D$ is represented by the matrix
\be \label{eq:T.nonunit}
  {\BbbT} =  \pmatrix{ 1 & 0 & 0 &0 \cr  {t'}_1 & \lambda_1 & 0 & 0
  \cr {t'}_2 & 0 & \lambda_1 & 0 \cr  {t'}_3 & 0&  0 & \lambda_3 }
\ee
The vector ${\bf t'} = ({t'}_1, {t'}_2, {t'}_3)$ is equal to
$R_2 {R_1}^{T} {\bf  t}$ in the notation of  (\ref{eq:T.SVD.nonpos}).
In this case, the image of the set of pure state density matrices
$\rho = \half[I + \bw \dtsig ]$ (with $|\bw| = 1$) under the action of
$\Phi_D$ is the translated ellipsoid
\be \label{eq:ellip.nonunit}
 \left( \frac{ x_1 - {t'}_1}{\lambda_1} \right)^2 +
   \left( \frac{ x_2 - {t'}_2}{\lambda_2} \right)^2 +
   \left( \frac{ x_3 - {t'}_3}{\lambda_3} \right)^2 = 1,
\ee
and again the image under $\Phi$ is a rotation of this.

\bigskip
It will be useful to write out explicitly the action of
the diagonal unital map
$\Phi[{\lambda}_1, {\lambda}_2, {\lambda}_3]$ on a density matrix
in the form
\be \label{Phi}
\Phi[{\lambda}_1, {\lambda}_2, {\lambda}_3](\rho)
& = &  \Phi[{\lambda}_1, {\lambda}_2, {\lambda}_3]
\pmatrix{a & b \cr b^{\dagger} & c}
  \\ & = & \nonumber
\half \pmatrix{ (a+c) + {\lambda}_{3}(a-c) & {\lambda}_1(b+b^{\dagger})
+ {\lambda}_2(b-b^{\dagger}) \cr {\lambda}_1(b+b^{\dagger}) -
{\lambda}_2(b-b^{\dagger}) &
(a+c)- {\lambda}_{3}(a-c)}.
\ee

\subsection{Complete Positivity Conditions}
\label{sect:singv.cond}

\bigskip

The requirement that $\Phi$ be stochastic imposes a number of
constraints on the matrix ${\BbbT}$.  We describe these in
Appendix B in which we give explicit formulas for the matrix
elements of ${\BbbT}$ in terms of the Stokes parameterization of
the operators $A_k$. These formulas in turn imply constraints
on the eigenvalues $({\lambda}_1, {\lambda}_2, {\lambda}_3)$
described in the previous section.

Let $T_{jk}$ denote the elements of $\BbbT$ using the convention
that $j,k \in 0\ldots 3$. Then the point with coordinates
$(T_{11}, T_{22}, T_{33}) $ must lie inside a tetrahedron
with corners at
$(1,1,1), (1,-1,-1), (-1,1,-1), (-1,-1,1)$.
These conditions are equivalent to four linear inequalities
which can be written compactly as
\begin{eqnarray}\label{eq:fuji.dcond}
|T_{11} \pm  T_{22} |  & \leq & |1 \pm T_{33}|.
\end{eqnarray}
(Note that we always have $T_{00} = 1$.)

In the special case where $\Phi$ is unital,
(\ref{eq:fuji.dcond}) implies
that the eigenvalues (which are necessarily real) satisfy
\begin{eqnarray}\label{eq:fuji.cond}
|\lambda_1 \pm {\lambda}_2|  & \leq & |1 \pm {\lambda}_3| .
\end{eqnarray}
In fact, for unital $\Phi$ the condition
(\ref{eq:fuji.cond}) is a necessary and sufficient condition
for the numbers $({\lambda}_1, {\lambda}_2, {\lambda}_3)$ to arise
as eigenvalues of the self-adjoint part of a unital stochastic map.

These conditions were discussed earlier by Algoet and Fujiwara
\cite{AF}. In addition they gave conditions for complete positivity
of some non-unital maps. In particular, for the special case of
(\ref{eq:T.nonunit}) with the form
\be \label{eq:AF.nonunit}
\Phi\big( \half[I + \bw \dtsig]\big) = \half
 \left[I + w_1 \lambda_1 \sigma_1  + (t + w_3 \lambda_3) \sigma_3 \right],
\ee
they showed that the necessary and sufficient condition for
complete positivity is
\be \label{eq:cpmcond.nonunit}
\lambda_1^2 + t^2 \leq (1 - |\lambda_3 |)^2
\ee

\subsection{Relation between capacity and minimal entropy for unital maps}
\label{subsect:prelim.cap}

We now assume wlog that $\Phi$ is self-adjoint and written in the
diagonal form $\Phi[{\lambda}_1, {\lambda}_2, {\lambda}_3]$.
Let $\mu = \max
(|{\lambda}_1|, |{\lambda}_2|, |{\lambda}_3|)$,  and let
${\bw}_{\mu}$ be a unit vector satisfying $T{\bw}_{\mu}= \pm \mu
{\bw}_{\mu}$. Then it is easy to show that
\be
\inf_{\rho} S[\Phi(\rho)] =
S\Big( \Phi \big( \half [I + {\bw}_{\mu} \dtsig] \big) \Big) =
h( \mu  )
\ee
where
\be
h( \mu ) = - \half (1 + \mu)  \ln \half (1 + \mu)
- \half (1 - \mu)  \ln \half (1 - \mu).
\ee

Consider now the question of computing the Holevo capacity  $C_{\rm Hol}(\Phi)$
defined in (\ref{Holevo}).
The choice ${\rho}_1 = \half [I + {\bw}_{\mu} \dtsig]$ and
${\rho}_2 = \half [I - {\bw}_{\mu} \dtsig]$, and ${\pi}_1 = {\pi}_2 =
\half$, both maximizes the first term $S(\rho) = S(\half I) =
\ln 2$ and minimizes the second term $\sum {\pi}_i S({\rho}_i) =
h(\mu)$. Hence it also maximizes their difference, which gives
$C_{\rm Hol}(\Phi) = \ln2 - h(\mu)$.

\medskip
In Section \ref{sect:summary} we stated several corollaries of
Conjecture \ref{conj:minent}. Here we will show how these
corollaries follow from
the assumption that minimal entropy is additive for unital maps.
\medskip

So suppose that Conjecture \ref{conj:minent} holds for unital maps, that is
suppose that the minimum value of $S(\Phi \otimes \Phi({\rho}_{12}))$ over
states
${\rho}_{12}$ on ${\bf C}^{2 \times 2}$ is $2 h(\mu)$. Then
there are four product states, namely ${\rho}_i = \half [I \pm {\bw}_{\mu}
\dtsig]
\otimes \half [I \pm {\bw}_{\mu} \dtsig]$, such that $S(\Phi \otimes
\Phi({\rho}_{i}))$ achieves this minimum value for each $i$. If we take
${\pi}_i = 1/4$ for each $i$, then $\rho = 1/4 I \otimes I$ and
$S(\Phi \otimes \Phi(\rho)) = \ln 4$ achieves its maximum possible value.
Hence with these choices, we can separately maximize each term on the right
side of (\ref{Holevo}) and therefore maximize the Holevo capacity. Therefore,
for unital maps the equality $C_{\rm Hol}(\Phi \otimes \Phi) =
2 C_{\rm Hol}(\Phi)$ is implied by the minimal entropy conjecture.
This demonstrates Corollary 2. Furthermore,
the two minimal entropy states $\half [I \pm {\bw}_{\mu} \dtsig]$ are
orthogonal. Hence both $C_{\rm Hol}(\Phi)$ and
$C_{\rm Hol}(\Phi \otimes \Phi)$ are achieved with orthogonal states,
and this establishes Corollary 3. Also, a simple calculation shows that
the expression inside the $\sup$ in the definition of
$C_{\shan}(\Phi)$ in (\ref{Shannon}) equals
$C_{\rm Hol}(\Phi)$ when we choose the input states to be
${\rho}_{i} = \half [I \pm {\bw}_{\mu} \dtsig]$ with ${\pi}_{i}=\half$, and
the POVM to be $E_{i} = \half [I \pm {\bw}_{\mu} \dtsig]$. This shows the
first statement of Corollary 4, and the second statement follows
immediately. Then Corollary 5 is a direct consequence.

\medskip

The essential observation in this argument is that we can find a
partition of unity in terms of a set of
{\it orthogonal} input states which are mapped into a level set of
minimal entropy.  For such inputs,
uniform averaging  yields the state $\rho = I$ whose output
$\Phi(\rho) = I$ has maximal entropy.  Hence
 both terms in the Holevo capacity are simultaneously maximised.
On ${\bf C}^2$ orthogonal inputs have the form
$\half [I \pm {\bw} \dtsig]$, and the corresponding output states
have the same entropy if and only if $\Phi$ is unital.
In that case, the products of these states form a set of
orthogonal inputs on ${\bf C}^4$ which map onto a level set of
entropy under the product map $\Phi \otimes \Phi$.   If the
minimal entropy is additive, one such set of product states
will be mapped onto a set of minimal entropy.
 For {\it non-unital} maps on ${\bf C}^2$, and more general
(non-product) maps on ${\bf C}^3$ or ${\bf C}^4$, this
need not hold.   (Fuchs and Shor \cite{Shor} have found an explicit
example of a map on ${\bf C}^3$ which does not have such a set
of orthogonal inputs.)   Hence the
above argument is quite special and does not provide a direct
link between the additivity of minimal entropy and additivity
of the Holevo capacity.

\section{Upper Bound on Norm} \label{sect:norm.bnd}

For any linear map $\Omega$  define
\be \label{eq:phi.norm}
M_{\Omega} \equiv \sup_{\rho \in {\DM}} \|\Omega(\rho)\|
  = \sup_{Q > 0} \frac{\|\Omega(Q)\|}{\tr Q}
\ee
so that for any $\varrho > 0$,  $\|\Omega(\varrho)\| \leq
M_{\Omega} \tr \varrho$.  Since the matrix norm  $\| ~ \cdot ~\| $
used in (\ref{eq:phi.norm}) is convex,
it suffices to consider the supremum over pure states or,
equivalently one-dimensional projections $\rho = |\psi \kb \psi|$.
Then $M_{\Omega}$ can be rewritten using the representation
(\ref{eq:kraus})
\be
 M_{\Omega}  = \sup_{\psi, \chi} \sum_k
      \left| \bra \chi , A_k  \psi \ket \right|^2
\ee
where the supremum is taken over those vectors
satisfying $|\psi| = |\chi| = 1$.

We restrict attention now to unital maps.
As discussed in Section \ref{sect:notation}, wlog we assume
that $\Phi$ is diagonal of the form
$\Phi[{\lambda}_1, {\lambda}_2, {\lambda}_3]$.
Then it follows from the discussion in section \ref{sect:notation}
that
\be
M_{\Phi} = \half \big(1 + {\max}_k|{\lambda}_k|\big) .
\ee

In this section
we will show  that for unital maps on
${\bf C}^{2 \times 2}$, the norm  $M_{\Phi}$ is multiplicative, i.e.,
$M_{\Phi \otimes \Omega} = M_{\Phi} M_{\Omega}.$  In fact, we will
show a slightly stronger result.
\begin{thm} \label{thm:normbnd}
 Let $\Omega$ be any 2-positive map on
${\bf C}^{n \times n}$ and let
$\Phi$ be a unital stochastic map on ${\bf C}^{2 \times 2}$.
Then $M_{\Phi \otimes \Omega} = M_{\Phi} M_{\Omega}$.
\end{thm}

Notice that Theorem \ref{thm:normbnd}  implies that
$||(\Phi \otimes \Omega)({\rho})||$ is maximised on product
states of the form $\rho = {\rho}_{1} \otimes {\rho}_2$,
where $||\Phi({\rho}_1)|| = M_{\Phi}$ and $||\Omega({\rho}_2)|| = M_{\Omega}$.

\medskip

Our proof will need the following
well-known result.  (See, e.g., \cite{HJ1}.)
\begin{lemma} \label{lemma:poscond}
Let  $S = \pmatrix{A & B \cr B^{\dagger} & C}$ be a matrix in block form
with $A, C > 0$ .
Then $S$ is (strictly) positive definite (i.e. $S > 0$) if and only if
$ A > B C^{-1} B^{\dagger}$ if and only if $C > B^{\dagger} A^{-1} B$.
\end{lemma}
As immediate corollaries we find
\begin{cor}
Let  $S = \pmatrix{A & B \cr B^{\dagger} & C}$ be a matrix in block form
with $A, C$ positive semi-definite.  Then $S$ is positive semi-definite
if and only if for all $u > 0$ one of the following
 two equivalent conditions holds
\bee
A + uI & > &  B(C +uI)^{-1} B^{\dagger}  \\
C + uI & > &  B^{\dagger}(A +uI)^{-1} B .
\eee
\end{cor}
\begin{cor} \label{cor:schwarz}
If  $S = \pmatrix{A & B \cr B^{\dagger} & C} \geq 0$, then
\be \label{eq:block.ineq}
\| B \|^2 = \|B B^{\dagger}\| \leq \|A\| \, \|C\|.
\ee
\end{cor}
To prove this note that
\bee
\bra v, BB^{\dagger} v \ket & \leq &
   \| C + u I \| \, \bra v, B(C+uI)^{-1}B^{\dagger} v \ket \\
  & < & \| C + u I \| \, \bra v, (A+uI) v)   \ket
  ~ \leq  ~ \| C + u I \|  \, \| A + u I \| \, \|v\|^2.
\eee
Choosing $v$ an eigenvector of $BB^{\dagger}$ and letting $u \raw 0$
proves (\ref{eq:block.ineq}).

\bigskip

Returning to the proof of Theorem \ref{thm:normbnd},
let $\rho =  \pmatrix{\rho_1 & \gamma \cr \gamma^{\dagger} & \rho_2}$
be a density matrix on ${\bf C}^2 \otimes {\bf C}^n$ written in
block form with
$\rho_1, \rho_2, \gamma$ each $n \times n$ matrices.
 First observe that
the two-positivity of $\Omega$ implies that
\be
\Big( I \otimes \Omega \Big) (\rho) =
 \pmatrix{\Omega(\rho_1) & \Omega(\gamma) \cr
    \Omega(\gamma)^{\dagger} & \Omega(\rho_2)} \geq 0
\ee
Hence, it follows from (\ref{eq:block.ineq}) that
\be \label{ineq:bnd1}
 \| \Omega(\gamma) \|^2 = \| \Omega(\gamma) \Omega(\gamma)^{\dagger} \|
 &  \leq & \| \Omega(\rho_1) \|\, \| \Omega(\rho_2) \| \nonumber \\
  & \leq & \tr \rho_1 \, \tr \rho_2 \, M_\Omega^2
\ee

Now use the form  (\ref{Phi}) and the linearity
 of $\Omega$ to write
\be
\lefteqn{ \Big( \Phi \otimes \Omega \Big) (\rho) =
\pmatrix{ P & L \cr L^{\dagger} & Q }} \\
& = & \nonumber \half
 \pmatrix{\Omega[\rho_1 + \rho_2 + \lambda_3(\rho_1 -\rho_2) ] &
(\lambda_1 + \lambda_2) \Omega(\gamma) +
   (\lambda_1 - \lambda_2) \Omega(\gamma)^{\dagger}
  \cr (\lambda_1 + \lambda_2) \Omega(\gamma)^{\dagger} +
   (\lambda_1 - \lambda_2) \Omega(\gamma)  &
\Omega[\rho_1 + \rho_2 - \lambda_3(\rho_1 -\rho_2) ]}
\ee
Note that  the complete positivity of $\Phi$ implies that
$\rho_1 + \rho_2 + \lambda_3(\rho_1 -\rho_2) > 0$ .  Thus
if $x = \tr \rho_1$
\be
 \| P \| & \leq & \nonumber
 M_{\Omega}\, \half \tr[ \rho_1 + \rho_2 + \lambda_3(\rho_1  -\rho_2)] \\
   & = &M_{\Omega} \Big[ \half + \lambda_3 (x - \half) \Big],  ~~~
\hbox{and} \label{eq:Pnorm}
\\  \label{eq:Qnorm}
 \| Q \| & \leq & M_{\Omega} \Big[ \half - \lambda_3 (x - \half)  \Big].
\ee

Now we can assume wlog that $\lambda_3 = \max_k |\lambda_k|$ so
that $M_{\Phi} = \half (1+ \lambda_3)$.
Then to prove Theorem \ref{thm:normbnd}, it suffices to show that
\be \label{eq:zcond}
 z > \half (1+ \lambda_3) M_{\Omega}  \imp
   zI - \Big( \Phi \otimes \Omega \Big) (\rho) > 0.
\ee
Note that
\be \label{ineq: Lnorm}
\| L L^{\dagger} \| \leq  (z - \|P\|)(z - \|Q\|)
\ee
and the general property $P \leq \|P \|$ imply
\bee
L (zI - P)^{-1} L^{\dagger} & \leq & L  (z - \|P\|)^{-1} L^{\dagger}
    \leq \| L L^{\dagger} \| (z - \|P\|)^{-1} \\
& \leq & (z - \|Q\|)  \leq zI - Q.
\eee
Therefore,  by Lemma \ref{lemma:poscond}, to verify (\ref{eq:zcond}),
 it suffices to show (\ref{ineq: Lnorm}). But  it follows from
(\ref{ineq:bnd1})   that
\bee
\| L L^{\dagger} \|  & = &  \frth  \Big\|  \Big[
 (\lambda_1+\lambda_2) \Omega(\gamma) + (\lambda_1-\lambda_2)
\Omega(\gamma^{\dagger}) \Big] \Big[
 (\lambda_1+\lambda_2) \Omega(\gamma)^{\dagger} + (\lambda_1-\lambda_2)
\Omega(\gamma) \Big]  \Big\| \\
& \leq & \frth \left[(\lambda_1+\lambda_2)^2 + 2
  |\lambda_1+\lambda_2| \, |\lambda_1-\lambda_2| + (\lambda_1-\lambda_2)^2
\right] \| \Omega(\gamma) \|^2 \\
& \leq & \lambda_1^2 x(1-x) M_{\Omega}^2.
\eee
where we have used $\| \Omega(\gamma) \|^2  = \| \Omega(\gamma)^{\dagger} \|^2
 = \| \Omega(\gamma) \Omega(\gamma)^{\dagger} \| $.
However, (\ref{eq:Pnorm}) and (\ref{eq:Qnorm}) also imply
\bee
(zI - \|P\|) (zI - \|Q\|)\geq \Big (\lambda_3(1-x) M_{\Omega} \Big)
     \Big (x \lambda_3M_{\Omega} \Big) =
   x(1-x) \lambda_3^2  M_{\Omega}^2.
\eee
Since we have assumed $ \lambda_3^2 >  \lambda_1^2$, these inequalities
imply  (\ref{ineq: Lnorm}).

\bigskip

\noindent{\bf Remark:}  At this point the {\em only} use we made of the
unital character of $\Phi$ was
\nl (a) to give a specific formula for $M_\Phi$
\nl (b) to use the special form (\ref{Phi}) of representing $\Phi$.

It is possible to generalize these formulas to some non-unital
stochastic maps. Any stochastic map $\Phi$ can be written in the form
(\ref{eq:T.nonunit}),  where $\BbbT$ is the $4 \times 4$ matrix which
represents its action on $(I, {\sigma}_{1}, {\sigma}_{2}, {\sigma}_{3})$.
Suppose $\lambda_1$ is the smallest diagonal entry of $\BbbT$.
If $t_1=0$, then the above method can be extended
in a straightforward way to deduce that
$\Phi$ also satisfies the conclusion of Theorem \ref{thm:normbnd},
for any values of $t_2$, $t_3$ allowed by complete positivity.
That is, as long as we do not translate the ellipsoid in the
direction of its shortest major axis, the conclusion
still holds. This is consistent with the conclusions of
Section \ref{sect:non.unit}, where we argue
that this is the hardest case to analyse. The difficulty
occurs when the two other major axes have equal lengths,
so that the ellipsoid is a `flying saucer'. This produces
a circle of states of maximal norm and minimal entropy in the ellipsoid.
It is necessary to show that no entanglement of these minimal
entropy states can increase the norm above the product bound,
or can lower the entropy below the product sum. We discuss this
situation further in (\ref{subsect:nonu.min.ent}).

\section{Minimal Entropy Analysis} \label{sect:entanal}

\subsection{Reduction via Convexity} \label{sect:convex}

\begin{lemma}
Let $\Phi, \Omega$ be unital stochastic maps with
$M_{\Phi}$ and $M_{\Omega}$ equal to $\mu$
and $\nu$, respectively.  Then
\be
  \inf_{\rho~ \pure} S(\Phi \otimes \Omega)(\rho) \geq
    \inf_{\rho ~\pure} S\Big(\Phi[\mu,u,\mu] \otimes
 \Omega[\nu,v,\nu]\Big)(\rho)
\ee
where $|u| \leq \mu$ and $|v| \leq \nu.$
\end{lemma}

 \bigskip

\noindent{\bf Proof of Lemma:} By the results of sections
\ref{sect:notation} and \ref{sect:singv.cond}, we can assume
wlog that $\Phi$ and
$\Omega$ are self-adjoint and diagonal, with eigenvalues
$(\lambda_1, \lambda_2, \lambda_3)$ and $(\omega_1, \omega_2, \omega_3)$
respectively, where $\lambda_3 = \mu > 0$ and
$\omega_3 = \nu > 0.$
We first consider the case $\mu, \nu > 1/3$.
 It  follows
from (\ref{eq:fuji.cond}) that the eigenvalues
$\lambda_1 , \lambda_2$ lie in a convex set with extreme points
\bee
(\mu,\mu),~ (\mu,2\mu-1),~ (2\mu-1, \mu),~ (-\mu,-\mu),~ (-\mu,1-2\mu),
 ~  (1-2\mu, -\mu)
\eee
If we let
 $\Phi_1 \equiv \Phi[\mu,\mu,\mu],~ \Phi_2 \equiv \Phi[\mu,2\mu-1,\mu]$
etc. so that $\Phi_j ~~ (j=1 \ldots 6)$ denote the stochastic maps
corresponding to these six points, we can write
$\Phi = \sum_{j=1}^6 a_j \Phi_j$ as a convex combination of these six maps
and similarly for $\Omega = \sum_{j=1}^6 b_k \Omega_k$.  Then, since the
entropy is concave we find
\be \label{eq:conv.prearg}
S(\Phi \otimes \Omega)(\rho) & = &
  S\left( \Big[\sum_{j=1}^6 a_j \Phi_j \Big] \otimes
 \Big[\sum_{k=1}^6 b_k \Omega_k \Big]\right)(\rho)  \nonumber \\
 & = & S \left( \sum_j \sum_k  a_j b_k \Phi_j  \otimes \Omega_k \right)
     (\rho) \nonumber \\
  & \geq &  \sum_j \sum_k  a_j b_k S \Big( \Phi_j  \otimes \Omega_k
    \Big)  (\rho) \\   \nonumber
   & \geq &  \min \{ S \Big( \Phi_j  \otimes \Omega_k \Big)  (\rho) :
    j = 1 \ldots 6,~  k=1\ldots 6 \}
\ee
But now we note that
$\Phi_4 = {\Upsilon}_{3} \circ \Phi_1$ and $\Phi_5 = {\Upsilon}_{3} \circ
\Phi_2$, where ${\Upsilon}_{3} (\rho) = \sigma_z \rho \sigma_z$.
Hence, e.g.,
\bee
   (\Phi_5  \otimes \Omega_4 )(\rho) =
   (\sigma_z \otimes \sigma_z) \Big[ (\Phi_2  \otimes \Omega_1 )(\rho)
   \Big]  (\sigma_z \otimes \sigma_z)
\eee
so that
\bee
   S \Big( \Phi_5  \otimes \Omega_4 \Big )(\rho) =
      S \Big( \Phi_2  \otimes \Omega_1 \Big )(\rho) .
\eee
and similarly for $S \Big( \Phi_1  \otimes \Omega_4 \Big )(\rho)
= S \Big( \Phi_1  \otimes \Omega_1 \Big )(\rho) $ etc.
Hence we can replace (\ref{eq:conv.prearg}) by
\be \label{eq:conv.arg}
S(\Phi \otimes \Omega)(\rho) \geq
    \min \{ S \Big( \Phi_j  \otimes \Omega_k \Big)  (\rho) :
    j,k=1,2,3 \}.
\ee
Since we also have
$ \inf_{\rho ~\pure} S\Big(\Phi[\mu,u,\mu] \ot \Omega (\rho)\Big) =
   \inf_{\rho ~\pure} S\Big(\Phi[u, \mu,\mu] \ot \Omega (\rho)\Big))$
for any $\Omega$, and
since $\Phi_j,~ j=1,2,3$ and $\Omega_k,~ k=1,2,3$ have the form
given in the lemma, the result follows.

For $\mu < 1/3$ we proceed similarly, but with the convex set
given by the rectangle with corners $(\pm \mu, \pm \mu)$.

\subsection{Special Form of Pure State} \label{sect:entropy}

As shown in Section \ref{sect:convex}, to show additivity of
minimal entropy for unital stochastic maps it is sufficient to
consider self-adjoint maps of the special form
$\Phi[\mu, u, \mu]$ and $\Omega[\nu, v,\nu]$. In this section
we prove additivity for these maps over a special class of entangled
states.

\begin{thm} \label{thm:main}
Let $\Phi[\mu, u, \mu]$ and $\Omega[\nu, v,\nu]$ be diagonal
stochastic maps, satisfying
$\mu \geq |u|$ and $\nu \geq |v|$, so that
$\mu$ and $\nu$  are the largest eigenvalues of $\Phi$ and $\Omega$
respectively. Let $| \psi \ket $ be a pure state
of the form
$| \psi \ket = a |0 \, 0 \ket +
    e^{i \theta} d |1 \: 1 \ket$. Then
\be
 S\big(\Phi \otimes \Omega\big) \big( | \psi \kb \psi | \big) \geq
   h(\mu) \,+\, h(\nu)
  = \inf_{\rho} S[\Phi(\rho)] \,+\, \inf_{\gamma} S[\Omega(\gamma)].
\ee
\end{thm}

We prove Theorem \ref{thm:main} in the next section. Here we derive
some intermediate results which will be used in the proof.
For generality we consider diagonal maps
$\Phi[{\lambda}_1, {\lambda}_2, {\lambda}_3]$ and
$\Omega[\omega_1, \omega_2,\omega_3]$, and for definiteness we
also assume that
$|\lambda_3|$ and $|\omega_3|$ are their largest singular
values.

\bigskip

We find the density
matrix for a pure state of  the form
$ | \psi \ket = a |00 \ket + e^{i \theta} d | 11 \ket $
with $a, d$ real  and $a^2 + d^2 = 1$.
Then if $\rho = | \psi \kb \psi | $
\begin{eqnarray} \label{eq:rho.diag}
\rho & = & \left( \begin{array}{cccc}
  a^2 & 0 & 0 & ad e^{i \theta} \\
  0 & 0 & 0 & 0 \\
  0 & 0 & 0 & 0 \\
 ad e^{-i \theta} & 0 & 0 & d^2
\end{array} \right) =
 \left( \begin{array}{cccc}
  \alpha & 0 & 0 & e^{i \theta} \sqrt{t}/2 \\
  0 & 0 & 0 & 0 \\
  0 & 0 & 0 & 0 \\
e^{-i \theta} \sqrt{t}/2 & 0 & 0 & 1 - \alpha
\end{array} \right) \\
 & = & \frth \alpha ~ \nonumber
  [ I \ot I + \sigma_z \ot \sigma_z + I \ot \sigma_z +  \sigma_z \ot I ] \\
 & ~ & + \frth (1 - \alpha )~
  [ I \ot I + \sigma_z \ot \sigma_z - I \ot \sigma_z - \sigma_z \ot I ]\\
 & ~ & + \frth \sqrt{t} ~ \big[ \cos \theta \,
  (\sigma_x \ot \sigma_x - \sigma_y \ot \sigma_y )
 - \sin\theta \,  (\sigma_x \ot \sigma_y + \sigma_y \ot \sigma_x )
    \big] \nonumber
 \end{eqnarray}
where $\alpha = a^2$ and $t = 4 \alpha (1 - \alpha)$, so that
$t \in [0,1]$.
Applying the stochastic maps gives
\begin{eqnarray}
[\Phi \ot \Omega] (\rho)
 & = & \frth \alpha ~  [ I \ot I + \lambda_3 \omega_3 \sigma_z \ot
\sigma_z + \omega_3 I \ot \sigma_z +  \lambda_3 \sigma_z \ot I ] \\
 & ~ & + \frth (1 - \alpha )~   [ I \ot I + \lambda_3 \omega_3 \sigma_z \ot
\sigma_z - \omega_3 I \ot  \sigma_z - \lambda_3 \sigma_z \ot I ]
\nonumber \\
 & ~ & + \frth \sqrt{t}  ~ \big[ \cos \theta \,
~(\lambda_1 \omega_1 \sigma_x \ot \sigma_x - \lambda_2 \omega_2 \sigma_y
\ot \sigma_y)  \nonumber \\
& ~ & \quad\quad\quad\quad - \sin\theta \,
(\lambda_1
\omega_2 \sigma_x \ot \sigma_y +
\lambda_2 \omega_1 \sigma_y \ot \sigma_x ).    \big] \nonumber
\end{eqnarray}
Notice that because $\Phi$ and $\Omega$ are diagonal, the result remains a
linear combination of terms of the form
$\sigma_k \ot \sigma_k ~~ (k = 0 \ldots 3)$ and
$I \ot \sigma_z$ and $\sigma_z \ot I$; no cross terms of the form
$\sigma_x \ot \sigma_z$ etc arise.  Therefore, the only non-zero
terms in $[\Phi \ot \Omega] (\rho)$ lie along the diagonal or
skew diagonal.  Thus  $ [\Phi \ot \Omega](\rho)$ can be written
in the form
\begin{eqnarray*}
 \pmatrix{ X & 0 & 0 & X \cr 0 & X & X & 0 \cr
    0 & X & X & 0 \cr X & 0 & 0 & X }
\end{eqnarray*}
where $X$ denotes a non-zero matrix element. Thus,
$[\Phi \ot \Omega](\rho)$ is equivalent to a block
diagonal matrix.

A straightforward computation shows that these blocks, which
we refer to as ``outer'' and ``inner'' can be written respectively as
\begin{eqnarray}\label{block:out}
\frth  \left( \begin{array}{cc}
 ~ 1 + \lambda_3 \omega_3 +  \sqrt{1-t}(\lambda_3 + \omega_3)  ~ &
         ~ \half \sqrt{t}(e^{i \theta} \lambda_+ \omega_+ + e^{- i \theta}
\lambda_{-} \omega_{-}) ~ \\
  \half \sqrt{t}(e^{i \theta} \lambda_{-} \omega_{-} + e^{- i \theta}
\lambda_{+} \omega_{+}) &
 1 + \lambda_3 \omega_3 - \sqrt{1-t}(\lambda_3 + \omega_3)
\end{array} \right)
\end{eqnarray}
and
\begin{eqnarray}\label{block:in}
\frth \left( \begin{array}{cc}
   1 - \lambda_3 \omega_3 + \sqrt{1-t}(\lambda_3 - \omega_3)  &
   \half \sqrt{t}(e^{i \theta} \lambda_+ \omega_{-} + e^{- i \theta}
\lambda_{-} \omega_{+})  \\
   \half \sqrt{t}(e^{i \theta} \lambda_{-} \omega_+ + e^{- i \theta}
\lambda_{+} \omega_{-}) &
  1 - \lambda_3 \omega_3 - \sqrt{1-t}(\lambda_3 - \omega_3)
\end{array} \right) .
\end{eqnarray}
where $\lambda_{\pm} = \lambda_1 \pm \lambda_2$ and similarly for
$\omega_{\pm}$.

The first has eigenvalues
\be\label{eq:eval.out}
\frth
\Big[ 1 + \lambda_3 \omega_3\Big]
 \pm \frth\Big[ { (1-t)(\lambda_3 + \omega_3)^2
  +  \frth t(\lambda_{+}^2 \omega_{+}^2 + \lambda_{-}^2 \omega_{-}^2
+ 2 \cos (2 \theta) \gamma) } ~\Big]^{1/2}
\ee
while the second has eigenvalues
\be\label{eq:eval.in}
\frth \Big[ 1 - \lambda_3 \omega_3\Big]
\pm \frth\Big[ { (1-t)(\lambda_3 - \omega_3)^2
  +  \frth t(\lambda_{+}^2 \omega_{-}^2 + \lambda_{-}^2 \omega_{+}^2
+ 2 \cos (2 \theta) \gamma) } ~\Big]^{1/2}
\ee
where $\gamma = \lambda_{+} \lambda_{-} \omega_{+} \omega_{-}$.

Minimum entropy occurs when both pairs of eigenvalues are spread out as far as
possible. This happens either at $\theta=0$ (if $\gamma \geq 0$)
or at $\theta=\pi/2$ (if $\gamma \leq 0$).
We will be interested in the case $\lambda_1 \geq |\lambda_2|$ and
$\omega_1 \geq |\omega_2|$, which means that $\gamma \geq 0$,
so we assume that $\theta=0$ henceforth.

We need more compact notation for the eigenvalues. Define
\be\label{def:f(t)}
f(t) = \Big((\lambda_3 + \omega_3)^2 - t [(\lambda_3 + \omega_3)^2
-(\lambda_1 \omega_1 + \lambda_2 \omega_2)^2] \Big)^{1/2}
\ee
and
\be\label{def:g(t)}
g(t) = \Big((\lambda_3 - \omega_3)^2 - t [(\lambda_3 - \omega_3)^2
-(\lambda_1 \omega_1 - \lambda_2 \omega_2)^2] \Big)^{1/2}
\ee
Also define
\be
A = 1 + \lambda_3 \omega_3, \quad\quad
B = 1 - \lambda_3 \omega_3
\ee
Then the four eigenvalues of $[\Phi \ot \Omega](\rho)$ at $\theta=0$ are simply
\be
\frth [A \pm f(t)], \quad \quad \frth [B \pm g(t)]
\ee

The state $| \psi \ket $ is unentangled when $t=0$ and maximally
entangled for $t=1$. We want
to show that the entropy is minimized at $t=0$.
We will let $S(t)$ denote the entropy of $[\Phi \ot \Omega](\rho)$
considered as a function of $t.$
To analyze its behavior,
it is convenient to use the function
\be
  {\eta}(\alpha,x) = - (\alpha+x) \log(\alpha+x) - (\alpha-x)
\log(\alpha-x).
\ee
It follows that
\be
  S(t) = \frth  {\eta}[A, f(t)]  +   \frth  {\eta}[B, g(t)]   + \log 4,
\ee
where we have used the fact that $\half(A+B) = 1$.  To find the minimum
of $S(t)$ it suffices to analyze the behavior of $ {\eta}[A, f(t)]$ and
$ {\eta}[B, g(t)].$  First observe that
\be \label{eq:1st.deriv}
  \frac{d~}{dt} {\eta}[A, f(t)] =  f^{\prime}(t)
        \log  \frac{ A- f(t)}{A+f(t)}
\ee
and
\be \label{eq:2nd.deriv}
 \frac{d^2~}{dt^2} {\eta}[A, f(t)] & = & f^{\prime\prime}(t)
   \log  \frac{ A- f(t)}{A+f(t)} - |f^{\prime}(t)|^2
   \frac{2A}{A - [f(t)]^2} \nonumber \\
  & \leq &   \frac{2A}{A - [f(t)]^2}
  \big[ f^{\prime\prime}(t) f(t) - |f^{\prime}(t)|^2 \big]
\ee
if $f^{\prime\prime}(t) \leq 0$ and $0 \leq f(t) \leq A$.
This follows from the elementary inequality
$ \log \left( \frac{1+x}{1-x} \right) \geq \frac{2x}{1-x^2}$
(which holds for $ x \in [0,1]$) applied to $x = f(t)/A$.
Now $f(t)$ is a function of the form $\sqrt{a - bt}$ for which one
easily checks that $f^{\prime\prime}(t) < 0$ and
$|f^{\prime\prime}(t)| f(t) - |f^{\prime}(t)|^2 = 0$.
Therefore, it follows immediately from (\ref{eq:2nd.deriv}) that
$\frac{d^2~}{dt^2} {\eta}[A, f(t)] \leq 0 $.  Since $g(t)$ also
has the form $\sqrt{a - bt}$, a similar argument
holds for $ {\eta}[B, g(t)].$
Hence $S^{\prime\prime}(t) < 0$ from which we
conclude that $S(t)$ is a concave function on $[0,1]$, and therefore
 attains its minimum at either $t=0$ or $t=1$.

Hence to prove that $S$ attains its minimal value at $t=0$,
it is necessary and sufficient to show that $S(1) \geq S(0)$.
In essence, we have  shown that for a state of the form
given in Theorem 1 {\em the minimal entropy is attained
for either a maximally entangled state or a simple product
state.}

If we again think of $f(t)$ in the form $\sqrt{a - bt}$, then
(\ref{def:f(t)}) (together with our assumption that $\lambda_3$
and $\omega_3$ are the maximal singular values)
implies that $b > 0$. Combined with
(\ref{eq:1st.deriv}) this implies that ${\eta}[A, f(t)] $ is increasing.
However, this need not be true for $g(t)$.  For example, when
$\lambda_3 = \omega_3$ and $t=1$,
$g^{\prime}(1) = \half \, |\lambda_1 \omega_1 - \lambda_2 \omega_2|$
which implies that $g(t)$ is increasing and ${\eta}[B, g(t)] $
decreasing.   Thus, the general situation is that the
entropy $S(t)$ is a linear combination of two concave functions
corresponding to the contributions from the ``outer'' and ``inner''
eigenvalues respectively.  The former is always increasing,
while the latter can be decreasing as $t$ goes from $0$ to $1$.
To illustrate this, Figure \ref{fig:eigenmove} shows how
the eigenvalues of the product state move in the case where
${\lambda}_3={\omega}_3=\mu$. The ``inner'' eigenvalues are
both $\frth(1-{\mu}^2)$, and they move apart as $t$ increases
away from $0$, which lowers their contribution to the entropy.
The ``outer'' eigenvalues $\frth(1\pm{\mu})^2$ move
together as $t$ increases, which raises their contribution
to the entropy.

To examine the difference $S(1) - S(0)$, we note that
\be
 S(1) & = & \log 4 +
  \frth {\eta}(1 + \lambda_3 \omega_3, |\lambda_1 \omega_1 + \lambda_2
\omega_2|) + \frth {\eta}(1 - \lambda_3 \omega_3, |\lambda_1 \omega_1 -
\lambda_2 \omega_2|)
 ~~~ \\ S(0) & = & \log 4 +
  \frth {\eta}(1 + \lambda_3 \omega_3, \lambda_3 + \omega_3)
~~~~~~ + \frth {\eta}(1 - \lambda_3 \omega_3, |\lambda_3  - \omega_3|)
\ee
so that
\be \label{eq:entdiff}
4[ S(1) - S(0)] & = &
  {\eta}(1 + \lambda_3 \omega_3, |\lambda_1 \omega_1 + \lambda_2 \omega_2|)
  -    {\eta}(1 + \lambda_3 \omega_3, \lambda_3 + \omega_3) \\
& ~ & +  {\eta}(1 - \lambda_3 \omega_3,
    |\lambda_1 \omega_1 - \lambda_2 \omega_2|)
-   {\eta}(1 - \lambda_3 \omega_3, |\lambda_3  - \omega_3|). \nonumber
\ee
Since $\frac{\partial ~}{\partial x} {\eta}(\alpha,x) =
  \log \frac{\alpha-x}{\alpha+x} < 0 $ if $x > 0$, ${\eta}(\alpha,x)$ is
decreasing in $x$.  By our assumptions,
$  \lambda_3 \geq |\lambda_1 | \geq |\lambda_1 \omega_1 | $ and
$  \omega_3 \geq |\omega_2 | \geq |\lambda_2 \omega_2 | $
so that
\bee
  \lambda_3 + \omega_3 >  |\lambda_1 \omega_1 + \lambda_2 \omega_2|,
\eee
and hence the difference of the the first two terms in (\ref{eq:entdiff})
(which corresponds to the change in entropy from the ``outer'' eigenvalues)
is always positive.  The change from the inner eigenvalues need not
be positive however; indeed, when $\lambda_3  = \omega_3$ it must be
negative.  Thus we need to show that the contribution from the
inner eigenvalues cannot dominate.

We gain some intuition from an elementary analysis of ${\eta}(\half, x)$.
This function is largest near $x = 0$, where it is flat, but has its
largest derivative near $x = \pm 1$.  Hence one expects the change
from the larger `` outer'' eigenvalues to dominate.  Explicit analysis
of the extreme points in the next section verifies this.

\bigskip

For the proof of Theorem \ref{thm:main} we will restrict to the
values $\lambda_1 = \lambda_3 = \mu$ and $\lambda_2 = u$ with $|u| \leq
\mu$, and
$\omega_1 = \omega_3 = \nu$ and $\omega_2 = v$ with $|v| \leq \nu$.
In this case (\ref{eq:entdiff}) becomes
\be \label{eq:entdiffspec}
4[ S(1) - S(0)] & = &
  {\eta}(1 + \mu \nu, \mu \nu  + u v)
  -    {\eta}(1 + \mu \nu, \mu + \nu) \\
& ~ & +  {\eta}(1 - \mu \nu,
    \mu \nu - u v)
-   {\eta}(1 - \mu \nu, \mu - \nu). \nonumber
\ee

\subsection{Analysis of Extreme Points}

In this section we will complete the proof of Theorem \ref{thm:main}.
By the argument in section \ref{sect:convex}, it suffices to
consider either
$u = \pm \mu $ with  $\mu \in [0,\thrd]$, or $u = \mu $ with
$\mu \in [\thrd,1]$, or $u = 2\mu-1$
with $\mu \in [\thrd, 1]$.  Similarly for $\Omega$: either
$v = \pm \nu$ with $\nu\in [0,\thrd]$, or $v = \nu$
with $\nu \in [\thrd, 1]$, or $v = 2 \nu-1$
with $\nu \in [\thrd, 1]$.
So we wish to prove the positivity of $S(1) - S(0)$ for these values of
the parameters.

The simplest case occurs when $u = \mu$, $v = \nu$.
In this case  (\ref{eq:entdiffspec}) becomes simply
\be \label{eq:entdiffspec2}
4[ S(1) - S(0)] & = &
  {\eta}(1 + \mu \nu, 2 \mu \nu)
  -    {\eta}(1 + \mu \nu, \mu + \nu) \\
& ~ & +  {\eta}(1 - \mu \nu,
    0)
-   {\eta}(1 - \mu \nu, \mu - \nu). \nonumber
\ee
Since ${\eta}(\alpha, x)$ is decreasing in $|x|$, the first term
dominates the second, and the third term dominates the fourth,
hence $ S(1) - S(0) > 0$ in this case.

The remaining cases are handled numerically. It is useful to first consider
a special case, namely $\mu = \nu$.
This arises when the convexity argument is applied to the product
channel $\Phi \ot \Phi$, which is the situation of most interest to us.
Then (\ref{eq:entdiffspec2}) yields
\be \label{eq:entdiff:Phi=Omega}
4[ S(1) - S(0)] & = &
  - (1 + 2{\mu}^2  + uv) \log (1 + 2{\mu}^2 + uv) \nonumber \\
& ~ & - (1 - 2{\mu}^2 + uv) \log (1 - 2{\mu}^2 + uv)
 - 2(1 - uv) \log (1 - uv) \nonumber \\
& ~ & + 4(1 + {\mu}) \log (1 + {\mu}) +
4(1 - {\mu}) \log (1 - { \mu}).
\ee

Graphing verifies that (\ref{eq:entdiff:Phi=Omega}) is positive for
the two extreme values $uv = \pm {\mu}^2$ in the range $0 \leq {\mu} \leq
\thrd$ (see Figure \ref{fig:entdiff1}),
and for the three extreme values $uv = {\mu}^2$, $uv = {\mu}(2{\mu}-1)$
and $uv=(2{\mu}-1)^2$ in the range $\thrd \leq {\mu} \leq 1$
(see Figure \ref{fig:entdiff2}).
The graphs show that (\ref{eq:entdiff:Phi=Omega}) is
smallest when $\mu \simeq 0,1$, so we analyze these regions more
carefully. In the first case when $0 \leq {\mu} \leq \thrd$ and $uv = -
{\mu}^2$,
we expand around  $\mu = 0$. This gives
\be \label{eq:entdiff:x=0}
4[ S(1) - S(0)]  \simeq
  4 {\mu}^2 \nonumber
\ee
In the second case when $\thrd \leq {\mu} \leq 1$ and $uv = {\mu}(2{\mu}-1)$,
write $x = 1 -\mu$, and expand in $x$:
\begin{eqnarray}
4S(1)-4S(0) & \simeq & 3x \log {1 \over x} + x
\Big[7(1+\log 4)-6 \log 3 -4(1 + \log 2)\Big]\nonumber\\
& \simeq & 3x \log {1 \over x} + 3.34 x.
\end{eqnarray}
This is manifestly positive for $x$ small. Similarly in the case
$uv=(2{\mu}-1)^2$ we have
\begin{eqnarray}
4S(1)-4S(0) & \simeq & 4x\log {1 \over x} + 4 x (1-\log 2)\nonumber\\
& \simeq & 4x\log {1 \over x} + 1.23 x
\end{eqnarray}

\bigskip
The general case $\mu \neq \nu$ is handled similarly.
We have
\be \label{eq:entdiff:Phi!=Omega}
4[ S(1) - S(0)] & = &
  - (1 + 2 {\mu} \nu  + uv) \log (1 + 2 {\mu} \nu + uv) \nonumber\\
& ~ & - (1 - 2 {\mu} \nu + uv) \log (1 - 2 {\mu} \nu + uv)
 - 2(1 - uv) \log (1 - uv) \nonumber\\
& ~ & + 2(1 + {\mu}) \log (1 + {\mu}) +
2(1 - {\mu}) \log (1 - { \mu}) \nonumber\\
& ~ & + 2(1 + {\nu}) \log (1 + {\nu}) +
2(1 - {\nu}) \log (1 - { \nu}).
\ee

By symmetry it suffices to assume that $\nu \leq \mu$. For
$0 \leq \mu \leq \thrd$ and $0 \leq \nu \leq \mu$ we have two
extreme values $uv = \pm \mu \nu$.
Graphing (\ref{eq:entdiff:Phi!=Omega}) shows that it is positive
in both of these cases. Again the smallest values occur near
$\mu = \nu =0$, so we expand (\ref{eq:entdiff:Phi!=Omega}) around
this point. For  both values $uv = \pm \mu \nu$ this gives
\be \label{eq:entdiff:Phi!=Omega:case1}
4[ S(1) - S(0)]  \simeq
  2 ({\mu}^2 + {\nu}^2). \nonumber
\ee
For $\thrd \leq \mu \leq 1$ and $0 \leq \nu \leq \thrd$,
there are four extreme values $uv = \mu \nu$, $uv = - \mu \nu$,
$uv = (2 \mu -1) \nu$ and $uv = - (2 \mu -1) \nu$.
The graph of (\ref{eq:entdiff:Phi!=Omega}) is positive in all cases, with
smallest values around $\mu = \thrd$.
For $\thrd \leq \mu \leq 1$ and $\thrd \leq \nu \leq \mu$,
there are also four extreme values, $uv = \mu \nu$,
$uv = {\mu}(2{\nu}-1)$, $uv = (2{\mu}-1){\nu}$ and
$uv = (2{\mu}-1)(2{\nu}-1)$
(see Figure \ref{fig:entdiff3} for the last of these).
The graphs of (\ref{eq:entdiff:Phi!=Omega}) are positive in each case,
and the smallest values occur near $\mu=\nu=1$. This region can be
analyzed more carefully by expanding the functions to leading order in
$x=1-{\mu}$ and $y=1-{\nu}$. For example, when
$uv = (2{\mu}-1)(2{\nu}-1)$ the expansion of (\ref{eq:entdiff:Phi!=Omega})
yields
\begin{eqnarray}\label{approx}
4S(1)-4S(0) & \simeq & 2 [x \log x + y \log y - 2(x+y) \log (x+y)]
\nonumber \\
& & + x(2 + 2 \log 2) + y(2 + 2 \log 2)
\end{eqnarray}
Using convexity of the function $x \log x$, we can bound (\ref{approx})
from below by $2(x + y)$, which demonstrates positivity for
$x,y$ small. Similar results are obtained for the other cases.

\subsection{Mixing Discussion} \label{sect:mix}

In this section we extend Theorem \ref{thm:main} to pure states
$|\psi \ket$ formed from any
entanglement of states of minimal
entropy.  After a precise statement of this and proof of this
extension, we discuss its interpretation and the evidence
for more general validity of Conjecture \ref{conj:minent}.

Let $\Phi$ be a unital stochastic map. As explained in
Appendix A, we can write $\Phi = U {\Phi}_{S} U^\dg$ where
$U$ is a unitary operator and ${\Phi}_{S}$ is self-adjoint.
Let $\mu = ||S||$.  Define
\be\label{def:L}
{\cal L}(\Phi) = \{ \rho = \half (I + N) \,|\, {\Phi}_{S}(N) = \pm \mu
N \}.
\ee
In words, ${\cal L}(\Phi)$ is the collection of density matrices which lie
in the direction of the largest eigenvalue of ${\Phi}_{S}$. If this largest
eigenvalue is non-degenerate, then ${\cal L}(\Phi)$
is a line segment between antipodal points on the Bloch sphere. In case of
degeneracy it may
be a disk, or even the entire Bloch sphere.

If ${\rho}_{12}$ is a density matrix on ${\bf C}^2 \ot {\bf C}^2$,
we denote by
${\rho}_1 = T_2({\rho}_{12})$ and ${\rho}_2 = T_1({\rho}_{12})$
the reduced density matrices on
${\bf C}^2$ obtained by taking the indicated partial traces.
\bigskip

\begin{thm} \label{thm:mixing}
Let $\Phi$ and $\Omega$ be unital
stochastic maps.
Let ${\rho}_{12}$ be a density matrix on
${\bf C}^2 \ot {\bf C}^2$, such that ${\rho}_1$ lies in
${\cal L}(\Phi)$, and ${\rho}_2$ lies in
${\cal L}(\Omega)$. Then
\be
 S\big(\Phi \otimes \Omega\big) \big( {\rho}_{12} \big) \geq
   \inf_{\rho} S[\Phi(\rho)] \, +\, \inf_{\gamma} S[\Omega(\gamma)].
\ee
\end{thm}

\noindent {\bf Proof}:
Wwe assume wlog that $\Phi$ and $\Omega$ are diagonal maps in the form
${\Phi}[ {\lambda}_1, {\lambda}_2, {\lambda}_3] $ and
${\Omega}[ {\omega}_1, {\omega}_2, {\omega}_3] $. Furthermore we
can arrange that ${\lambda}_3 = \mu$ is the largest eigenvalue,
so that $\half [ I \pm {\sigma}_3]$
lies in ${\cal L}(\Phi)$. Similarly we can arrange that
$\omega_3 = \nu$ is the largest eigenvalue, so that
$\half [ I  \pm {\sigma}_3]$ also
lies in ${\cal L}(\Omega)$.

Wlog we can assume that ${\rho}_{12} = | \psi \kb \psi |$ is a pure state.
In the bases
which diagonalize $\Phi$ and $\Omega$, we have
\be
| \psi \ket = a_{00} | 00 \ket + a_{01} | 01 \ket + a_{10} | 10 \ket +
a_{11} | 11 \ket
\ee
Define the matrix $A$ to be
\be
A = \left(\matrix{a_{00} & a_{01} \cr a_{10} & a_{11}\cr}\right)
\ee
Then as shown in Appendix A, the reduced density matrices are
\be
{\rho}_1 = A A^{\dagger}, \qquad\qquad
{\rho}_2 = \Big( A^{\dagger} A \Big)^{\rm T}
\ee
We obtain the ``Schmidt decomposition'' by applying the singular
value decomposition to $A$. The result is a new basis in which $| \psi \ket$
has the diagonal form assumed in Theorem \ref{thm:main}.
This decomposition is obtained by finding unitary operators
$U_1, U_2$ which diagonalize $A A^{\dagger}$ and $A^{\dagger} A$
respectively, so that $U_1 A U_2^{\dagger}$ is also diagonal. By assumption
${\rho}_1$ lies in ${\cal L}(\Phi)$, and hence so does
$A A^{\dagger}$. If $\mu$ is non-degenerate, then ${\cal L}(\Phi)$
is the line segment consisting of the diagonal density matrices.
Therefore
$A A^{\dagger}$ is also diagonal, so $U_1$ is equal to the identity,
up to a phase.
If $\nu$ is also non-degenerate then $U_2$ is also proportional to
the
identity, and hence ${\rho}_{12}$ is already in diagonal form. The
result follows immediately by applying Theorem \ref{thm:main}.

In general either or both of $\mu$ and $\nu$ may be degenerate.
For example, if $\mu$ is 2-fold degenerate, then ${\cal L}(\Phi)$ is a disk
containing the $z$-axis in the Bloch sphere. So $A A^{\dagger}$ lies in this
disk, and hence it is diagonalized by a rotation of the Bloch sphere which
preserves this disk.  By definition, such a rotation commutes with the action
of $\Phi$, since the plane which contains this disk is an eigenspace of
$\Phi$. Hence the unitary operator $U_1$ commutes with $\Phi$.
Similarly if $\mu$ is 3-fold
degenerate, then every unitary operator commutes with $\Phi$.
To apply the argument to $\Omega$, note that by assumption ${\rho}_2$ lies
in ${\cal L}(\Omega)$. If $\nu$ is non-degenerate, or is 3-fold
degenerate, the same argument applies. If $\nu$ is 2-fold degenerate,
then ${\cal L}(\Omega)$ is a disk.
The transpose operation on the Bloch sphere
is the reflection in the xz-plane, and this does not in general
preserve a disk containing the z-axis. However we have assumed that
$\Omega$ is diagonal, and hence ${\cal L}(\Omega)$ either lies in
the xz-plane or the yz-plane. In both cases the transpose leaves
${\cal L}(\Omega)$ invariant, and hence the same argument can be applied
to deduce that $U_2$ also commutes with
$\Omega$. It follows that $(\Phi \otimes \Omega)(| \psi \kb \psi |)$ is
unitarily equivalent to $(\Phi \otimes \Omega)(| \psi' \kb \psi' |)$
where $| \psi' \ket$ has the form assumed in Theorem \ref{thm:main},
 and hence the result follows.

\bigskip
\bigskip
\noindent{\bf Remarks:}
\begin{enumerate}
\item
The set ${\cal L}(\Phi)$ contains the states of minimal entropy for $\Phi$.
Theorem \ref{thm:mixing} shows that by entangling the minimal entropy
states of the individual channels we cannot decrease the entropy of the product
channel. Since it seems unlikely that entangling states of higher
entropy will improve the situation, we present this as strong
evidence for our conjecture.

\item
We illustrate Theorem \ref{thm:mixing} in the case where
$\Phi[{\lambda}_1, {\lambda}_2, {\lambda}_3]$ is self-adjoint and diagonal,
with ${\lambda}_1 = {\lambda}_3$.
If the $ |\psi \ket $ is real, i.e., the matrix $A = a_{jk}$ is real, then
the $2 \times 2$ unitary matrices $U_1$ and $U_2$ which diagonalize
$A A^{\dagger}$ and $A^{\dagger} A$ can be chosen real and orthogonal,
in which case we emphasize this by writing them as $\co_1$ and $\co_2$.
Now suppose that our original orthogonal basis
$ |0\ket, |1\ket $ on ${\bf C}^2$ corresponds to the eigenvectors of
$\sigma_z$ so that the corresponding pure state projections are
$\half [I + \bw \cdot \sigma]$, with
$\bw = (0,0,1)$ corresponding to the ``North pole'' of a sphere.
Each unitary $2 \times 2$ matrix $U$ can be associated with a
real orthogonal $3 \times 3$ matrix, and the effect of $U$ on the
basis vectors corresponds to a rotation on the sphere or the
action of a real orthogonal $3 \times 3$ matrix on  $\bw.$
When the original $2 \times 2$ matrix is real orthogonal, the
corresponding rotation on the sphere reduces to a rotation
in the xz-plane.

If we now write the unital stochastic map $\Phi$ in the form
\bee
\Phi : \rho = \half {[I + \bw \cdot \sigma]}   \raw
   \Phi(\rho) = \half [I + T \bw \cdot \sigma] ,
\eee
then the change of basis
is equivalent to replacing $T$ by $\widehat{\co} T \widehat{\co}^{-1}$
where $\widehat{\co}$ denotes the $3 \times 3$ orthogonal matrix
associated with $\co$.  Thus, for
example, when
$\co = \pmatrix{ ~\cos \, \theta/2 & \sin \, \theta/2 \cr
   - \sin \, \theta/2 & \cos \, \theta/2 } $, ~~
$\widehat{\co} = \pmatrix{ ~\cos \theta & 0 & \sin \theta \cr
 0 & 1 & 0 \cr  - \sin \theta & 0 & \cos \theta }$.
If $T$ is diagonal with eigenvalues $\lambda_x = \lambda_z$, then
$\widehat{\co} T \widehat{\co}^{-1} = T.$  Hence every state
$ |\psi \ket $ with real coefficients can be diagonalized, and the
matrix $T$ is unchanged.

\item
Suppose that $\Phi$ has one very large
singular value and two small ones.  Then the unit sphere corresponding
to the set of density matrices is mapped into an ellipsoid
shaped like a football, and the states of minimal entropy lie at
the ends of the football (see Figure \ref{fig:ellip1}).   We can interpret Theorem
\ref{thm:main} as saying that entangling these minimal entropy states will
not decrease the entropy below the sum of the minimum entropies.

Now suppose that we always keep two  eigenvalues equal, but vary the
 parameters so that the ends of the football move in until it
becomes a sphere and then a ``flying saucer'' (see Figure \ref{fig:ellip2}).
The
ends of the football have moved in to states corresponding to maximal
entropy.  The minimal entropy states now form a great circle.
As we explained above, our special form for $\psi$ allows
a general entanglement of states
corresponding to these great circles of minimal entropy.
Yet even this more general entanglement does not decrease
the entropy below that of product states.

\item
The discussion above shows that, at least in the case of unital
maps,  if Conjecture \ref{conj:minent} does not hold, then the
 entanglements which use states of higher
entropy would achieve a lower entropy on the product space than
entanglements of states of minimal entropy.  In addition,
such entanglements would have to lower the entropy without
increasing the largest eigenvalue of $\Phi(\rho_{12})$ beyond
the product value given in Theorem \ref{thm:normbnd}.  We do
not find this plausible.

\end{enumerate}

\section{Non-unital Maps}\label{sect:non.unit}

In this section we give some heuristic evidence to support
Conjecture \ref{conj:minent}.  Before doing so, we illustrate
the differences between unital and non-unital maps by discussing
some of the properties of a special class of maps on
${\bf C}^{2 \times 2}$.

A non-unital map is one for which $\Phi(I) \neq I$.  This means
that it takes a randomly distributed alphabet to a non-random
distribution.  One would intuitively expect that the maximum
capacity would then be achieved for alphabets which are {\em not}
evenly distributed.  Although this is true classically, it need
not be true for quantum stochastic maps as shown by the example
below.

It follows from equation (\ref{eq:Trep}) that a unital stochastic
map defines a linear map on the subspace of matrices with trace
zero. However, a non-unital map is affine when restricted to this
subspace.

\subsection{Special subclass} \label{sect:nonunit.sub}
We now consider non-unital maps which correspond, in the
notation of Section \ref{sect:notation} to the matrix
\be  \label{eq:phi.fuchs}
  {\BbbT} =  \pmatrix{ 1 & 0 & 0 &0 \cr  0 & \lambda_1 & 0 & 0
  \cr 0 & 0 & 0 & 0 \cr  t & 0&  0 & \lambda_3 }
\ee
with $t \neq 0$.  This is easily seen to yield the map
\be \label{eq:spec.nonunit}
\Phi\big( \half[I + \bw \dtsig]\big) = \half
 \left[I + w_1 \lambda_1 \sigma_1  + (t + w_3 \lambda_3) \sigma_3 \right].
\ee
If $\lambda_1 = \frac{1}{\sqrt{3}}$ and $t = \lambda_3 = \thrd$,
this is equivalent to the ``splaying'' channel introduced by Fuchs
\cite{Fu1} to demonstrate that there exist stochastic maps for
which the Holevo capacity (\ref{Holevo}) is achieved only by
non-orthogonal states. The case $\lambda_3 = 0$ was considered in
\cite{LesRu} in a different context.
The case $\lambda_1 = 0$ is essentially classical, i.e., if
$\rho$ is restricted to the subset of states of the form
$\half[I \pm w \sigma_3]$, the action of $\phi$ is equivalent
to the action of the column stochastic matrix
$\half \pmatrix{1 + t + \lambda_3 & 1 - t + \lambda_3 \cr
   1 - t - \lambda_3 & 1 + t - \lambda_3}$
on the probability vector $\half \pmatrix{ 1 + w \cr 1 - w}.$

Since equality in (\ref{eq:AF.nonunit}) holds for
Fuchs example, it can be regarded as an extreme case.  Because $\lambda_2
= 0$ for this class of maps, they map the unit sphere of density matrices
into the ellipse in the x-z plane satisfying the equation
\be \label{eq:ellipse}
   \frac{w_1^2}{\lambda_1^2} +  \frac{(w_3 - t)^2}{\lambda_3^2} = 1
\ee
In the special cases,  $\lambda_1 = 0$ and $\lambda_3 = 0$, these
ellipses become vertical and horizontal line segments respectively.

\bigskip

In the classical case ($\lambda_1 = 0$) it is not hard to
show that the maximal capacity is
{\em never} achieved for $\pi = \half$.  One has only two
pure states $\rho_{\pm} = \half [I \pm \sigma_3]$
(which {\em are} orthogonal)
for which $S[\Phi_{\pm}(\rho)]$ is not identical.

\bigskip

By contrast, for the non-unital quantum case with
 ${\lambda_1} > {\lambda_3}$, it appears, in general, that maximal
capacity is achieved at $\pi = \half$ and with non-orthogonal
states.  Moreover, these non-orthogonal states need not
correspond to the minimal entropy states.   Some insight
into these observations can be obtained by looking at the
ellipse (\ref{eq:ellipse})  in Figure \ref{fig:Fuchs} corresponding to Fuchs
channel. (Fuchs \cite{Fu1}  showed explictly that non-orthogonal
states are required to achieve maximum capacity, and that
the maximal capacity achievable with orthogonal states occurs
for $\pi = \half$.)
The endpoints of the ellipse (denoted $A_{\pm}$) correspond to
$\Phi \big( \half [I \pm \sigma_1] \big) $ and have
entropy $h \big[ \half (1 +  2/3)\big]$ while the minimal
entropy states (denoted $C_{\pm}$) correspond to the images of
$\Phi \left( \half \Big[ I +
 \big( \pm \frac{\sqrt{3}}{2}, 0, \half \big) \dtsig \Big] \right)$
and have entropy $h \big[ \half (1 +  1/\sqrt{2})\big]$.
Note that this is the point at which the ellipse meets the
circle   $x^2 + z^2 = \half$, which is a  level set for
the entropy on the Bloch sphere.

\bigskip

   The states $\half [I \pm \sigma_1]$ are the only pair of
orthogonal states with identical entropy.  If one tries to
move from $A_+$ toward $C_+$ to lower the entropy from one
of a pair of orthogonal states, the other orthogonal state
must move along the ellipse away from $A_-$, down and closer
to the origin and  hence has a higher entropy than $A_-$.
Explicit computation shows that the entropy price paid by moving
away from $A_-$ is greater than that gained by moving
$A_+$ toward $C_+$ .   For any pair of states $\rho_i ~~ (i = 1,2)$,
the state $\tilde{\rho} = \pi \tilde{\rho}_1 + (1-\pi) \tilde{\rho}_2$
which occurs in (\ref{Holevo}) lies along the line segment between
 $\tilde{\rho}_1 $ and $\tilde{\rho}_2$.  For the points $A_{\pm}$
it is easy to see that the maximum capacity will occur for
$\pi = \half$.  If one does not require orthogonal states, then
simultaneously moving both $A_{\pm}$ toward $C_{\pm}$ to decrease
the entropy seems advantageous.  However, this will also decrease
the entropy of the convex combination $\tilde{\rho}$ which
{\em increases} the capacity.  Hence, maximal capacity does not occur
at the minimal entropy states but at states on the ellipse which
lie between $C_{\pm}$ and $A_{\pm}$.  (As long as the move is
symmetric, maximal capacity will occur at $\pi = \half$.
Symmetry suggests this is optimal, but that does not seem to have
been proved explicitly.)

   One expects similar behavior for any map $\Phi$ of the form
(\ref{eq:phi.fuchs}) for which $\lambda_1 > \lambda_3$.
When $\lambda_1 < \lambda_3$ the major axis of the ellipse
lies along the x-axis, there is only one state of minimal entropy
at the ``top'' of the ellipse, and one expects the channel to
behave more and more like a classical channel as the ratio
$\lambda_1/\lambda_3$ increases.

When $\lambda_3 = 0$, the ellipse becomes a horizontal line so that the
minimal entropy states and the endpoints of the ellipse
coincide.  Hence, the maximal capacity is again achieved
for orthogonal states.  In the limiting case
$t^2 + \lambda_1^2 = 1$, the endpoints of the ellipse lie
on the unit circle of pure states and hence, have entropy
zero.  However, the capacity does {\em not} achieve its
maximum value of $\log 2$ but instead has the value
$h(t)$.

   As was noted earlier, $\Phi$ always has a fixed point.
For channels of the form (\ref{eq:phi.fuchs}), this fixed point
is at $(0, 0, \frac{t}{1 - \lambda_3})$.  However, we have
been unable to attach any significance to the fixed point.
(For Fuchs channel, the fixed point is at $\half [C_+ + C_-]$;
however, this seems coincidental.)

\bigskip

At first glance, it might seem that the price paid for the
versatility of non-unital channels is too great.  If we
fix $\lambda_2$ and $\lambda_3$, then when $t \neq 0$
the requirement (\ref{eq:cpmcond.nonunit}) implies
that $\lambda_1$ must be smaller (i.e., ``noisier'') than for
the corresponding unital channel.  For example, using
Fuchs values $t = \lambda_3 = \thrd, ~ \lambda_2 = 0 $,
one finds that $\lambda_1 = \frac{1}{\sqrt{3}} =  \approx 0.577 $ is
optimal and corresponds to the least ``noisy'' direction.  However,
if $t = 0$, one could increase this to $\lambda_1 = 2/3 \approx 0.667$
with corresponding decrease in noise so that the minimal
entropy (which comes from the states $\half[I \pm \sigma_3]$)
is $h(0.667)$.  For the non-unital channels these same states
would yield an entropy of only  $h(0.577)$.  However, the
states $\half
\Big[ I + \big( \pm \frac{\sqrt{3}}{2}, 0, \half \big) \dtsig\ \Big] $
will emerge with entropy $h(0.707)$.  The decrease in eigenvalues
of the part of $\Phi$ corresponding to the restriction to matrices of
trace zero, is overcome by the contribution to the emerging state
of the non-unital part of the map.  We expect that this behavior is
generic for non-unital maps.

\subsection{Minimal entropy}\label{subsect:nonu.min.ent}

Recall the discussion in Section \ref{sect:notation}.
For unital maps,
the generic situation is that there are two states of minimal
entropy corresponding to the endpoints  of the major axis.
In the generic situation for a non-unital map,
unless ${\mathbf t}$ is perpendicular to the major axis, the
ellipsoid in (\ref{eq:ellip.nonunit}) will have {\em only one} state
of minimal entropy.  Hence {\em any} entanglement will
require mixing with a state of higher entropy.  In the case
of a map (such as the Fuchs map (\ref{eq:spec.nonunit}) with
$|\lambda_3| > |\lambda_1|$) for which the translation ${\mathbf t}$
 is perpendicular  to the (non-degenerate) major axis of the ellipsoid,
there will  be two states of minimal entropy.  However, because the
translated ellipsoid is not centered at the origin,
these two states will not be the images of two orthogonal states,
but rather the images of two non-orthogonal pure states corresponding
to vectors $| \psi_1 \ket$ and $| \psi_2 \ket$.
This suggests that the best candidate for a minimal entropy entangled state
should have the form
$\Psi = \sum_{jk} a_{jk} | \psi_j \otimes \psi_k \ket $.
However, after changing to  orthogonal bases and using the SVD
decomposition  (\ref{eq:schmidt}), this can be rewritten in the form
$\Psi = a | \chi_1 \otimes \chi_3 \ket + d  | \chi_2 \otimes \chi_4 \ket$
where  $ \bra \chi_{1} , \chi_{2} \ket = \bra \chi_{3} , \chi_{4} \ket = 0$
and, at most, only {\em one} state in each of the pairs $\chi_1,
\chi_2$ and $\chi_3, \chi_4$
can equal either $\psi_1$ or $\psi_2$.  Hence any entanglement
must include states which are mapped into states of higher entropy
under the action of $\Phi_S$.
Thus one expects states of minimal entropy under the action of
$\Phi_S \otimes \Phi_S$ to have the product form
$| \psi_j \otimes \psi_k \ket $.

In the case of two-fold degeneracy (of the form
 $|\lambda_j| = |\lambda_k|$), a shift orthogonal to both
major axes will yield  a circle of
states of minimal entropy; however, that circle will {\em not} correspond
to a ``great circle'' but rather a circle of constant latitude. Such a circle
never includes the image of two orthogonal states and hence the argument
above still holds. In the case of three-fold degeneracy, the image will
be a sphere and {\em any} shift will again yield only a single state of
minimal entropy.

Since the maps $\Phi$ and $\Phi_S$ differ only by rotations of the
Bloch sphere, the same conclusions hold for $\Phi$.
Moreover, our analysis suggests that for {\em any} pair of maps $\Phi$ and
$\Omega$,
the states which yield minimum entropy under $\Phi \otimes \Omega$
will be simple product states, regardless of whether one or both
are non-unital.

\section{Conclusion}\label{sect:conc}


\bigskip

Our main result in this paper is that channel capacity is
additive for unital channels, where {\it unital} means that
the channel maps the totally random input state (whose density matrix
is proportional to the identity) into itself (for example, every
self-adjoint channel is unital). Specifically, we present strong evidence
that both the Shannon capacity (no entanglement for input states
or output measurements) and the Holevo capacity (inputs unentangled,
but output measurements may be entangled) are additive over two uses of
a two-dimensional unital channel. This is the first result
that establishes additivity of channel capacity for a broad
class of quantum channels.

We show that the problem reduces to finding the states of
{\it minimal entropy} for two uses of the channel. If these
minimal entropy states are {\it product
states}, then this implies additivity of capacity. We then prove
that one of two possibilities occurs:
either these minimal entropy states are product states,
or else they are entangled states whose reduced density matrices
have greater than minimal entropy. We argue that the latter
case is very unlikely (numerical experiments confirm this),
and so conclude that the former is true. As further
supporting evidence, we prove that the maximal norm of
states which emerge from the channel is multiplicative over two
channel uses.

\bigskip

Our results rely heavily on the Stokes parametrization and
the properties of the  image of the Bloch sphere under
stochastic maps.  In order to extend them to higher dimensions,
we would need an effective parameterization of the subspace
of  matrices of trace zero in higher dimensions.  One can always
write a density matrix in ${\bf C}^{d \times d}$ as
$\rho = \frac{1}{d} [I + N]$ where $\tr N = 0$.  However,
for $d = 4$, we do not know what the analog of the Bloch
sphere looks like.  We know only that its boundary
corresponds to those $N$ with eigenvalues $+3, -1, -1, -1$
which is not the analogue of the surface of a sphere.
Without knowing the geometry of this region, we can hardly hope
to answer the important question of how it transforms under
maps of the form $\Phi \otimes \Phi$.   Thus, we have been
forced to use indirect methods to reach conclusions about
the states of minimal entropy emerging from $\Phi \otimes \Phi$.

\bigskip


Our results have implications for the design of communication
channels.  Ideally, one wants to eliminate all noise.  However,
this will not be practical and one wants to know how best to
allocate resources.

In the case of unital channels, minimal entropy and maximal
capacity are achieved if signal codes are chosen to correspond
to the least noisy  ``direction'' or polarization.  (Here, we use
``direction'' in the sense of the Bloch sphere or maximum $\lambda_k$
in our notation.  This is unrelated to   direction of signal
transmission.) Hence, if maximizing capacity is the primary goal, then
it would seem sufficient to minimize noise in only one direction.
Even if the orthogonal directions are extremely noisy, signals
sent using optimal codes will not be affected.  However, in this
case ``classical communication'' becomes truly classical.  If
codes are restricted to one direction, then one is back in the
classical situation with one choice for encoding $0$ and $1$.
One has effectively lost the versatility of rotating the code basis
as a tool for such purposes as signal encryption.

Non-unital channels have far more versatility, some aspects of
which were discussed briefly in section \ref{sect:nonunit.sub}.
Much more work needs to be done analyzing the properties of
non-unital channels.  Thus far, most authors have looked for
examples of particular maps  which
illustrate particular facets of stochastic maps
(such as Fuchs \cite{Fu1} example demonstrating the possibility
of maximizing capacity with non-orthogonal states).  Our approach
has been to try to find parameters which characterize subclasses
of stochastic maps with certain properties.  As summarized in
Appendix C, most of the
known examples of noisy channels can easily be shown to belong
to one of the groups discussed in this paper.  A complete
analysis of non-unital maps would seem to require an
extension of conditions of the type (\ref{eq:AF.nonunit})
to general maps of the form (\ref{eq:T.nonunit}) with
$t_k, \lambda_k \neq 0$.

\appendix

\section{Singular Value, Polar and Schmidt Decompositions}

In this paper we make repeated, and sometimes subtle, use of the
singular value decomposition (SVD) of matrices.  In view of this,
and of some confusion in the literature about the connection
between the SVD and what is often referred to as the ``Schmidt''
decomposition, we provide a brief summary and review of
these closely related decompositions and their connection to
the better known polar decomposition.

We begin with the polar decomposition which can be extended to
bounded, and even some unbounded, operators \cite{RS1} on a Hilbert
space.
\begin{thm} \label{thm:polar}  {\em (Polar Decomposition)}
Any $m \times n$ matrix $A$ can be written in the form
$A = U |A| $ where the $n \times n$ matrix $|A| = \sqrt{A^{\dg} A}$ is
positive semi-definite and the $m \times n$ matrix $U$ is a partial
isometry.
\end{thm}
The term {\em partial isometry} means that $U^{\dg} U$ (or,
equivalently, $U U^{\dg}$) is a projection.  In general, $U$ need
not be unique but can be uniquely determined by the condition
$\ker U = \ker A$.  If $A$ is a square $n \times n$ matrix, then
$U$ can instead be chosen (non-uniquely) to be unitary.
Since $|A|$ is self-adjoint, it can be written as
$|A| = V D V^{\dg}$ where $D$ is a diagonal matrix with
non-negative entries and $V$ is unitary.  Inserting this
in Theorem \ref{thm:polar} with $U$ chosen to be unitary yields the SVD
since $W = UV$ is also unitary.
\begin{thm} \label{thm:SVD} {\em (Singular Value Decomposition)}
Any $n \times n$ matrix $A$ can be written in the form
$A = W D V^{\dg}$ with $V, W$ unitary and $D$ a positive
semi-definite diagonal matrix.
\end{thm}
The non-zero elements of $D$ are called the {\em singular values}
of $A$.  They are easily seen to be the eigenvalues of $|A|$ and,
hence, their squares yield the non-zero eigenvalues of $A^{\dg} A$.
As an immediate corollary, one finds that  $A^{\dg} A$ and
 $A A^{\dg}$ are unitarily equivalent and that $V$ and $W$ are,
respectively, the unitary transformations that diagonalize
$A^{\dg} A$ and  $A A^{\dg}$.  These
results can be extended to non-square matrices if the requirement that
$V,W$ be unitary is relaxed to partial isometry.

Using the notation of Section \ref{sect:notation},
we can apply the SVD to the $3 \times 3$
matrix $\rmT$ which corresponds to the restriction of the stochastic
map $\Phi$ to the subspace of matrices with trace zero.
Because $\rmT$ is real, the matrices $V,W$ can be chosen to be
real orthogonal so that we can write
\be\label{eq:T.SVD}
\rmT = {\cal  O}_1 \rmD {\cal  O}_2^{T}
\ee
where  ${\cal  O}_1$, ${\cal  O}_2$  are orthogonal and
the superscript $T$ denotes transpose. Now, every $3 \times 3$
 orthogonal matrix is either a rotation,
or the product of a rotation with the inversion $-I$.
Hence we can rewrite (\ref{eq:T.SVD}) as
\be\label{eq:T.SVD.nonpos}
\rmT = R_1 (\pm D) R_2^T = (R_1 R_2^T) R_2 (\pm D)  R_2^T
\ee
where $R_1$ and $R_2$ are rotations, and we conclude
that $\rmT$ can be written as
\be \label{eq:T.selfadj}
 \rmT =  R S
\ee
where $S$ is self-adjoint and $R$ is a rotation.
If $\Phi$ is unital, define the map $\Phi_S$ by
\be \label{eq:phi.selfadj}
\Phi_S\big( w_0 I + \bw \dtsig \big) =
w_0 I + S \bw \dtsig
\ee
Since every rotation is implemented by
a unitary transformation on ${\bf C}^2$, there is a unitary
operator $U$ such that for any state $\rho$,
\be
\Phi(\rho) = U \Phi_S ( \rho) U^{\dg}
\ee

For non-unital maps, a similar argument can be used to show
that any stochastic map has the form (\ref{eq:Trep})
where the restriction of $\Phi$ to the matrices with trace
zero has the form (\ref{eq:T.selfadj}) and
${\bf t} \raw R^T {\bf t} $, i.e.
$\Phi(\rho) = U \Phi_S ( \rho) U^{\dg}$ where
\be \label{Phi.selfadj}
\Phi_S\big(  w_0 I + \bw \dtsig \big) =
w_0 I + (R^T {\bf t} + S \bw ) \dtsig
\ee
and $R$ is the rotation on ${\bf R}^3$ corresponding to $U$.

\bigskip

By construction, either $S$ or $-S$ is positive semi-definite.
However, if $\Upsilon_k(\rho) = \sigma_k \rho \sigma_k$, then
 composing $\Upsilon_k$ with the diagonal map
$\Phi_D \Big( w_0 I +  \bw \dtsig  \Big) =
   w_0 I +  D \bw \dtsig $ merely changes the signs of
two of the diagonal elements of $D$.  Since $U \sigma_k$ is
also a unitary map, we can drop the restriction that $S$
be semi-definite by modifying $U$ if necessary.  This is useful
because, as we will see in the next section, the general conditions
on the eigenvalues of a matrix $S$ corresponding to a self-adjoint
map $\Phi$ include the possibility of negative and positive
eigenvalues.  We can summarize this discussion in the following
\begin{thm} \label{thm:polar.Phi}
Any stochastic map $\Phi$ on ${\bf C}^{2 \times 2}$ can be written
in the form $\Phi(\rho) = U \Phi_S ( \rho) U^{\dg}$
where $U$ is unitary and $\Phi_S$ is a stochastic map whose
 restriction to matrices with trace zero is self-adjoint.
\end{thm}

It may be worth noting that we can apply the polar decomposition
theorem directly to a completely positive map $\Phi$.
If we use  $\wh{\Phi}$ to denote the adjoint with respect to the
Hilbert-Schmidt inner product,  then Theorem \ref{thm:polar}
implies that we can write
\be \label{eq:polar.phi}
  \Phi = \Upsilon \circ |\Phi|
\ee
where $|\Phi| = \sqrt{ \wh{\Phi} \circ \Phi}$ and
$ \Upsilon$ is a partial isometry.   If $\Phi$ takes an algebra
(e.g., ${\bf C}^{n \times n}$) to itself, then $\Upsilon $ can be
chosen to be an isometry, i.e.,
${\wh \Upsilon } \Upsilon = \Upsilon  {\wh \Upsilon } = I$.
However, neither $|\Phi|$ nor $ \Upsilon$ need be stochastic
in general (even though their composition is).
On the contrary, if $\Phi_S$ has an odd number of negative
eigenvalues, then $\Phi_S = \Gamma \circ |\Phi| $ where
$\Gamma$ changes the sign of an odd number of eigenvalues
and, hence, is {\em not} a completely positive map.
(See Appendix C for further discussion and explicit examples.)

What we have shown in the argument above is that for unital
maps on ${\bf C}^{2 \times 2}$ , the isometry $\Upsilon$ can
always be implemented by a unitary transformation, possibly composed
with a map $\Gamma$ that takes
$[ w_0 I +  \bw \dtsig ] \raw [ w_0 I - \bw \dtsig ]$, i.e., there
is a unitary matrix $U$ such that  $\Upsilon$  has the form
$\Upsilon_{\pm}\Big( w_0 I +  \bw \dtsig  \Big) =
 U   [ w_0 I \pm  \bw \dtsig ]  U^{\dg}$ where one and only one
sign holds.  If $\Phi$ is a non-unital stochastic map, then $|\Phi|$
will not even be trace-preserving and the isometry  $\Upsilon$ will
correspond to a change of basis that mixes the identity $I$
with the three $\sigma$ matrices (in contrast to the unital case
in which the change of basis affects only the subspace of traceless
matrices spanned by the three Pauli matrices).
Hence for non-unital maps, the full polar decomposition
(\ref{eq:polar.phi}) of a
stochastic map $\Phi$ may be less useful than the polar
decomposition on the restriction to matrices of trace zero.
For unital maps on ${\bf C}^{n \times n}$ with $n \geq 3$ it
would be interesting to know how much $\Upsilon$ can differ
from a map of the form $\Upsilon(\rho) = U \rho U^{\dg}$
where $U$ is an $n \times n$ unitary matrix.

In order to see the connection between the SVD and the so-called
``Schmidt decomposition'', consider
a wave function or vector of the form
\be \label{eq:wavefctn}
   \Psi = \sum_{jk} a_{jk} \psi_j \otimes \chi_k
\ee
with $\{ \psi_j \}$  and $\{ \chi_k \}$ orthonormal.
It is not hard to see that there is an isomorphism between such
vectors and operators of the form
\be \label{def:Kpsi}
   K_\Psi = \sum_{jk} a_{jk}| \psi_j \kb \chi_k |
\ee
and that $ K_\Psi$ is a Hilbert-Schmidt operator if and only
if $\Psi$ is square-integrable (in the case of wave functions).
Moreover, if $\rho_{12} = |\Psi \kb \Psi |$, then
\bee
  \rho_1 \equiv T_2 (\rho_{12}) & = &  K_\Psi K_\Psi^{\dg} \\
 \rho_2 \equiv T_1 (\rho_{12}) & = &  (K_\Psi^{\dg} K_\Psi)^T
\eee
where $\rho_1$ and $\rho_2$ are the reduced density matrices
obtained by taking the indicated partial traces $T_2$ and $T_1$.
The ``Schmidt decomposition'' is an immediate
consequence of the application of the SVD to the
operator $K_{\psi}$ given by (\ref{def:Kpsi}) which implies
the following result.
\begin{thm}{\em (Schmidt)} Any wave function of the form
(\ref{eq:wavefctn}) can be rewritten as
\be \label{eq:schmidt}
   \Psi = \sum_{k} \mu_k \tilde{\psi}_k \otimes \tilde{\chi}_k
\ee
where $\mu_k$ are the singular values of the matrix $A$, the
bases $\{\tilde{\psi}_k \}$ and $\{\tilde{\chi}_k \}$ are
orthonormal and related by
$\mu_k \tilde{\psi}_k = K_\Psi \tilde{\chi}_k $ with $ K_\Psi$
given by (\ref{def:Kpsi}).
\end{thm}
It follows immediately that the reduced density matrices
$\rho_1$ and $\rho_2$ have the same non-zero eigenvalues
$\{{\mu_k}^2\}$
and $\{\tilde{\psi}_k \}$ and $\{\tilde{\chi}_k \}$ are
the eigenvectors of $\rho_1$ and $\rho_2$ respectively.

There is an interesting history associated with both the
SVD and Schmidt decompositions, as well as the attachment of Schmidt's
name to (\ref{eq:schmidt}) in the physics literature.
In Chapter 3 of \cite{HJ2}, Horn and Johnson give a detailed
account of the history of the SVD which goes back to Beltrami
and Jordan who independently obtained the SVD for real $n \times n$
matrices in the 1870's.   In 1902 Autonne obtained the SVD for
general nonsingular complex $n \times n$ matrices and later
made explicit the straightforward generalization to singular
matrices in a long paper in 1915 which seems to have been
subsequently overlooked by many researchers.

Independently, Schmidt obtained analogous results in 1907 for
operators associated with integral kernels.  When quantum chemists
became interested in density matrices in the 1960's, Carlson and
Keller \cite{CK} rediscovered some of his results.  However,
John Coleman \cite{Cole} soon pointed out the connection with
Schmidt's much earlier work.  Coleman's observation probably
made physicists and chemists aware of that work although he
did not use the term ``Schmidt decomposition'' which has
recently become popular in the quantum computing literature.
Physical chemists were initially interested in the reduced
density matrices which arise from a multi-particle wave function
of the form $\Psi(w_1 \ldots w_n)$ where $w_i$ denotes the space
and spin coordinates associated with the i-th particle.  It was
natural to decompose these coordinates into two subsets
$x = w_1 \ldots w_p$ and $y = w_{p+1} \ldots w_n$ so that
the density matrices $\rho_1$ and $\rho_2$ actually correspond to
 p-th and (N-p)-th order reduced density matrices of $\Psi$.
It is noteworthy that if the original wave function
$\Psi$ has some symmetry, then the functions
$\{\tilde{\psi}_k \}$ and $\{\tilde{\chi}_k \}$ can always
be chosen so that (\ref{eq:schmidt}) has the same symmetry.
  In particular, if $\Psi$ is antisymmetric, (as required by
the Pauli exclusion principle for fermions) then these
bases can be chosen so that (\ref{eq:schmidt}) is also antisymmetric;
i.e., there is no need to apply an additional antisymmetrizer.

In the case of multi-particle wave functions $\Psi(x,y)$
is readily interpreted as the kernel of
an integral  operator (which corresponds to $K_\Psi$ defined above)
acting on an infinite dimensional Hilbert space.
Hence, it is quite
natural to attribute the results to Schmidt when used in this context.
However, in quantum computation, one only considers the ``spin''
part of the wave function.  Since this can always be represented
by a finite dimensional matrix, the ``diagonalization'' of $a_{jk}$
in (\ref{eq:wavefctn}) can be obtained directly from the SVD;
there is no need to detour  into infinite dimensional Hilbert spaces
to use results for  integral operators.
Moreover, since the matrix form
of  the SVD preceded Schmidt's work,  it seems natural to use the term
SVD decomposition.

\section{Matrix Representation of Stochastic Maps}

To prove the eigenvalue conditions (\ref{eq:fuji.cond}), we  rewrite the
Kraus operators defined by (\ref{eq:kraus})  in the form
\be
A_k = v_{k0} I + {\bv}_k \dtsig
\ee
where $(v_{k0}, \bv_{k})$ is a vector in ${\bf C}^4$. We
will let $V$ denote the $n \times 4$  matrix with  elements $v_{kj}$,
$V_j$ its columns
as vectors in ${\bf C}^n$ and $\bV = (V_1, V_2, V_3)$.  Using the
relation
\be \label{eq:matprod}
(aI + \bu \dtsig )(bI + \bw \dtsig ) & = &
  (ab + \bu \cdot \bw) I + (a {\bw} + b {\bu} +
     i \bu \times \bw )\cdot \sigma
\ee
one finds
\bee
  A_{k} A_{k}^{\dagger} & = & \sum_{j=0}^3 |v_{kj}|^2 I +
  \big(v_{k0} \overline{\bv}_k + \overline{v}_{k0} \bv_k +
   i \bv_k \times \overline{\bv}_k \big) \dtsig  \\
 A_{k}^{\dagger} A_{k} & = & \sum_{j=0}^3 |v_{kj}|^2 I +
  \big(v_{k0} \overline{\bv}_k + \overline{v}_{k0} \bv_k -
   i \bv_k \times \overline{\bv}_k \big) \dtsig
\eee
so that
\be
\sum_k A_{k} A_{k}^{\dagger} & = & \big( \sum_{j=0}^3 |V_j|^2 \big) I +
2 \Big(  \Re  \bra V_0, \bV \ket +
  \sum_k \Re \bv_k \times \Im \bv_k \Big) \dtsig \\
\sum_k A_{k}^{\dagger} A_{k} & = & \big( \sum_{j=0}^3 |V_j|^2 \big) I +
2 \Big(  \Re  \bra V_0, \bV \ket -
  \sum_k \Re \bv_k \times \Im \bv_k \Big) \dtsig  \label{eq:Ident.form}
\ee
where $\la .,. \ra$ denotes the standard inner
product in ${\bf C}^n$ and $\bV = (V_1, V_2, V_3)$.
Hence if $\Phi$ is {\em either} unital or trace-preserving, then
\be \label{eq:Icond}
  \sum_{j=0}^3 |V_j|^2 = \tr \, V^{\dagger} V = 1.
\ee
In addition, it follows from (\ref{eq:cond.tracepres}) and
(\ref{eq:cond.unital}) that
\be  \label{eq:vcond.tracepres}
\tr \Phi(\rho) = \tr \rho & \imp &
 \Re  \bra V_0, \bV \ket +  \sum_k \Re \bv_k \times \Im \bv_k = 0 \\
\Phi(I) = I ~~ & \imp &
 \Re  \bra V_0, \bV \ket -  \sum_k \Re \bv_k \times \Im \bv_k = 0
  \label{eq:vcond.unital}.
\ee
This implies that $\Phi$ is both unital and trace-preserving if and
only if, in addition to (\ref{eq:Icond}),
\be
 \Re  \bra V_0, \bV \ket & = & 0   \label{iden2} \\
  \sum_k \Re \bv_k \times \Im \bv_k & = & 0 \label{iden3}
\ee
where (\ref{iden3}) can be rewritten as
\be\label{iden4}
 \Im \bra V_j, V_k \ket & = & 0, ~~~ j\neq k  \in 1,2,3
\ee
By defining $\wh{V}$ to be the matrix obtained by replacing
the first column of $V$ by $\wh{V}_0 = i V_0$, one can rewrite conditions
(\ref{iden2}) and (\ref{iden3}) as the single requirement that
that $\Im \, ( \wh{V}^{\dagger} \wh{V})$ is diagonal or, equivalently,
$ \Im  \bra \wh{V}_j, \wh{V}_k \ket = 0, ~~~ j\neq k \in 0 \ldots 3$.

\bigskip

We will now derive the general form of the real
$3 \times 3$ matrix $\rmT$ defined by (\ref{eq:T3rep}) so that
\be \label{eq:matrix.el}
{\rmT}_{jk} = \half {\rm Tr} \big({\sigma}_j \Phi({\sigma}_k)\big)
\ee
In general $\rmT$ is not symmetric; in fact, it is symmetric
if and only if every operator $A_k$ is self-adjoint.
One finds after straightforward calculation that
\be  \label{diags}
T_{jj} & = &
 \la V_0,V_0 \ra+ \la V_j,V_j \ra - \sum_{i \neq j} \la V_i, V_i \ra \\
\label{offdiags}
T_{ij} & = & 2 \Re \la V_i,V_j \ra \mp 2 \Im \la V_0, V_k \ra
\ee
where the $-$ holds in (\ref{offdiags}) if $\{i,j,k\}$ is an
even or cyclic
permutation of  $\{1,2,3\}$ and the $+$ sign if it is
an odd permutation.  Thus, for example,
\bee
T_{11} & = &
\la V_0,V_0 \ra + \la V_1,V_1 \ra - \la V_2, V_2 \ra - \la V_3,V_3 \ra \\
T_{12} & = & 2 \Re \la V_1,V_2 \ra - 2 \Im \la V_0,V_3 \ra
\eee

\bigskip

In the special case where $T$ is diagonal its eigenvalues can easily
be obtained from
(\ref{diags}). Let $({\lambda}_1, {\lambda}_2, {\lambda}_3)$ be the
eigenvalues, and define $q_j = \la V_j,V_j \ra ~~j=0 \ldots 3$.
Then (\ref{diags}) becomes
\be
{\lambda}_1 & = &  q_0 + q_1 -  q_2 - q_3  \nonumber \\
{\lambda}_2 & = & q_0 - q_1 + q_2 - q_3 \\
{\lambda}_3 & = & q_0 - q_1 -q_2 + q_3   \nonumber
\ee
Together with the condition (\ref{eq:Icond}) which can be written as
\be
1 & = &  q_0 + q_1 +  q_2 + q_3
\ee
this implies that the point
with coordinates
$({\lambda}_1, {\lambda}_2, {\lambda}_3)$ must lie inside the tetrahedron
with corners at
$(1,1,1), (1,-1,-1), (-1,1,-1), (-1,-1,1)$. Furthermore by taking $n \geq
4$ and
choosing the vectors $V_0, V_1, V_2, V_3$ to be orthogonal we see that
every point in this tetrahedron defines a triplet of eigenvalues which can
arise from a unital stochastic operator. These conditions are equivalent
to four linear inequalities
which must be satisfied by the eigenvalues, namely
\begin{eqnarray*}
{\lambda}_1 + {\lambda}_2  & \leq & 1 + {\lambda}_3 \\
{\lambda}_1  - {\lambda}_2  & \leq & 1 - {\lambda}_3 \\
 - {\lambda}_1 + {\lambda}_2  & \leq & 1 - {\lambda}_3 \\
- {\lambda}_1 -{\lambda}_2   & \leq & 1 + {\lambda}_3
\end{eqnarray*}
which is equivalent to the more compact (\ref{eq:fuji.cond}).

These conditions were obtained earlier by Algoet and Fujiwara
\cite{AF}.

\bigskip

The expression in (\ref{diags}) and (\ref{offdiags}) also
hold for non-unital $\Phi$.  To  calculate the non-zero elements in the
first column  extend (\ref{eq:matrix.el}) to $j = 0$ and
observe that  the trace-preserving condition
(\ref{eq:vcond.tracepres}) implies
$ \Re  \bra V_0, \bV_j \ket = -  \sum_k \Re \bv_k \times \Im \bv_k$.
Using this in (\ref{eq:Ident.form}) yields
\be
  t_j =T_{j0} = \half \tr \sigma_j \Phi(I) =
    4 \Re  \bra V_0, \bV_j \ket
\ee

\section{Examples}

We now give some examples of unital and non-unital maps
which illustrate some of the features discussed earlier
and show the correspondence between our parameterizations
and some well-known examples which are usually described
by their Kraus operators.  Following the notation of
Section \ref{sect:notation}, we will let
$\Phi[\lambda_1,\lambda_2,\lambda_3]$ denote a diagonal
unital map.

Before doing so we note that the classic example of a
linear, positivity preserving map which is {\em not} completely
positive is the transpose, which corresponds to
$\Phi[1, -1, 1]$.  We also note that if $\Upsilon_k$
denotes the maps $\Upsilon_k(\rho) = \sigma_k \rho \sigma_k$,
the  composition
$\Upsilon_k \circ \Phi[\lambda_1,\lambda_2,\lambda_3]$
changes the sign of the two eigenvalues whose subscript is
not $k$.  (For example
$\Upsilon_2 \circ \Phi[\lambda_1,\lambda_2,\lambda_3] =
\Phi[-\lambda_1,\lambda_2,-\lambda_3]  $.)  Thus the map
$\Phi[-1, -1, -1] $  which takes
$\half[I + \bw \dtsig] \raw \half[I - \bw \dtsig]$ is also
{\em not} completely positive.

\bigskip

\noindent {\bf Examples of Unital Channels}

 \begin{itemize}
\item[$\bullet$] Depolarizing channel:
$\Phi[1 - \frac{4x}{3}, 1 - \frac{4x}{3}, 1 - \frac{4x}{3}]$
\nl
\nl $ A_0 = \sqrt{1-x}\, I,
     ~~A_k = \sqrt{x/3} \, \sigma_k ~~ {(k=1,2,3)}$

\item[$\bullet$] BFS \cite{BFS} two-Pauli channel:
$\Phi[x, x, 2x -1]$
\nl
\nl $ A_0 = \sqrt{x} I, ~~A_k = \sqrt{\half(1-x)} \, \sigma_k,
   ~~ {(k=1,2)}$

\item[$\bullet$] Phase-damping channel:
$\Phi[1-x, 1-x, 1]$
\nl
\nl $ A_0 = \sqrt{1-x}\, I, ~~  A_{\pm} = \sqrt{x} \, \half[I \pm
\sigma_z]$

\item[$\bullet$]  Rotation: A single rotation is the simplest
example of a unital, non-self-adjoint map.
 $ A_1 = U$ where $U$ is unitary and $\det(U) = +1$.

A convex combination of such rotations using Kraus operators
 $ A_k = \tau_k U_k$ where
$\sum_k |\tau_k|^2 = 1$
yields a more general example of a unital map
which is not self-adjoint, namely,
$\Phi(\rho) = \sum_{k}
|\tau_k|^2 U_{k}^{\dagger} \rho U_{k}$.  Then the associated $3
\times 3$ matrix $\rmT = \sum_k  |\tau_k|^2 R_k$ where $R_k$ is
the rotation corresponding to $U_k$.
 \end {itemize}

The BFS two-Pauli channel was introduced by Bennett, Fuchs, and Smolin
\cite{BFS} to demonstrate that entangled states could reduce the
probability of error.  It is worth noting that this map corresponds
to the extreme points studied in detail in Section \ref{sect:entanal}.
 Hence, it is particularly noteworthy that our analysis
provides particularly strong evidence that entanglements do {\em not}
decrease the entropy for these maps.
   It is also worth noting that when $x = \thrd$ this map
becomes $\Phi[\thrd,\thrd, -\thrd]$.  Thus, although
$\Phi[1, 1, -1] $ and  $\Phi[-1, -1, -1] $ are {\em not}
completely positive, both
$\Phi[\thrd, \thrd, -\thrd] $ and  $\Phi[-\thrd, -\thrd, -\thrd] $
{\em are} completely positive.


\bigskip

\noindent{ \bf Examples of Non-Unital Channels}

 \begin{itemize}
\item[$\bullet$] Amplitude-damping channel
\nl $A_0 = \left( \begin{array}{cc}
  1 & 0  \\ 0 & \sqrt{1-t} \end{array}   \right), ~~
A_1 = \left( \begin{array}{cc}
  0 & \sqrt{t}  \\ 0 & 0 \end{array}   \right) $
\nl $ I  \raw  \Phi(I)  =
\left( \begin{array}{cc}  1+t & 0  \\ 0 & 1-t \end{array}   \right)
   = I + t \, \sigma_z$
\nl $~~~{\BbbT} = \left( \begin{array} {cccc}
 1 & 0 & 0 & 0 \\  0 &  \sqrt{1-t} & 0 & 0 \\
   0 &  0 & \sqrt{1-t}& 0 \\
 t &  0 & 0 & 1-t\end{array} \right) $
\nl Equality in
 $ ~~(\lambda_1 \pm{\lambda}_2)^2   \leq  (1 \pm {\lambda}_3)^2 - t^2  $

\item[$\bullet$] Fuchs  channel and
related examples discussed in
Section \ref{sect:nonunit.sub}

 \end{itemize}

\pagebreak

\pagebreak
\centerline{\underbar{Figure Captions}}

\begin{itemize}
\item[] {\bf Figure 1.} The four eigenvalues of the product state of minimal
entropy, with their direction of movement shown  as $u$ increases
away from $0$.

\item[] {\bf Figure 2.} The entropy difference $4[S(1) - S(0)]$ for the two
channels
\nl $\Phi[\mu, \mu, \mu] \otimes \Phi[\mu, \mu, \mu]$ (upper curve)
and $\Phi[\mu, \mu, \mu] \otimes \Phi[\mu, -\mu, \mu]$ (lower dashed curve),
in the range $0 \leq \mu \leq \thrd$. Note that the curves coincide
over most the interval.

\item[] {\bf Figure 3.}
The entropy difference $4[S(1) - S(0)]$ for the
three channels
\nl $\Phi[\mu, \mu, \mu] \otimes \Phi[\mu, \mu, \mu]$ (dots),
$\Phi[\mu, \mu, \mu] \otimes \Phi[\mu, 2\mu -1, \mu]$ (dashes),
and 
\nl $\Phi[\mu, 2\mu -1, \mu] \otimes \Phi[\mu, 2\mu -1, \mu]$ (full)
in the range $\thrd \leq \mu \leq 1$.

\item[] {\bf Figure 4.}
The entropy difference $4[S(1) - S(0)]$ for the channel
\nl $\Phi[\mu, 2\mu -1, \mu \otimes \Omega[\nu, 2\nu -1, \nu]$
in the range $\thrd \leq \mu, \nu \leq 1$.

\item[] {\bf Figure 5.}
The Bloch sphere and
its image (the ellipsoid) under a unital map with one large
and two small singular values. The endpoints of the ellipsoid
are shown separately -- these are the unique states of minimal entropy.

\item[] {\bf Figure 6.}
The Bloch sphere
and its image under a unital
map with two large and one small singular values. The
``waistband'' of the ellipsoid is shown separately;
this entire circle consists of minimal entropy states.

\item[] {\bf Figure 7.}
The Bloch sphere (the circle of
radius 1) and its image (the translated ellipse) under the
Fuchs map, together with the circle of radius
$1/\sqrt{2}$. The endpoints of the ellipse are marked $A
\pm$, and the points of minimal entropy are marked $C \pm$.
\end{itemize}

\pagebreak

\begin{figure}
\begin{center}
{\LARGE $ ~~~~~~ \,
 \curvearrowright \hspace{5.75cm} \curvearrowleft \, \curvearrowright
   ~~~~~~~~ \, \curvearrowleft   $} \newline
{\large $ 1\, \rule{2cm}{0.1mm}|\rule{7cm}{0.1mm}||
\rule{3.5cm}{0.1mm}|\rule{1.5cm}{0.1mm} \,0$ \newline $ ~~~~~~~
\frth(1 +|\mu|)^2 \hspace{5.0cm} \frth(1 - |\mu|^2)
   ~~~~~~~~~~ \frth(1 - |\mu|)^2 $}
\end{center}
\caption{The four eigenvalues of the product state of minimal
entropy, with their direction of movement shown  as $u$ increases
away from $0$. \label{fig:eigenmove}}
\end{figure}

\pagebreak

\begin{figure}
\centering{\resizebox{!}{3.0in}{\includegraphics{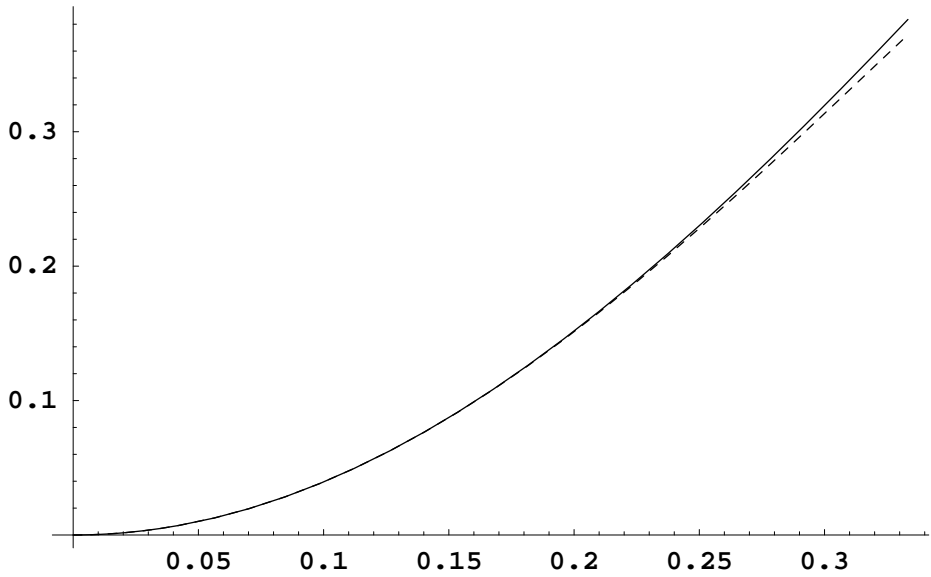}}}
\caption{The entropy difference $4[S(1) - S(0)]$ for the two
channels
\nl $\Phi[\mu, \mu, \mu] \otimes \Phi[\mu, \mu, \mu]$ (upper curve)
and $\Phi[\mu, \mu, \mu] \otimes \Phi[\mu, -\mu, \mu]$ (lower dashed curve),
in the range $0 \leq \mu \leq \thrd$. Note that the curves coincide
over most the interval.
\label{fig:entdiff1}}
\end{figure}

\pagebreak

\begin{figure}
\centering{\resizebox{!}{3.0in}{\includegraphics{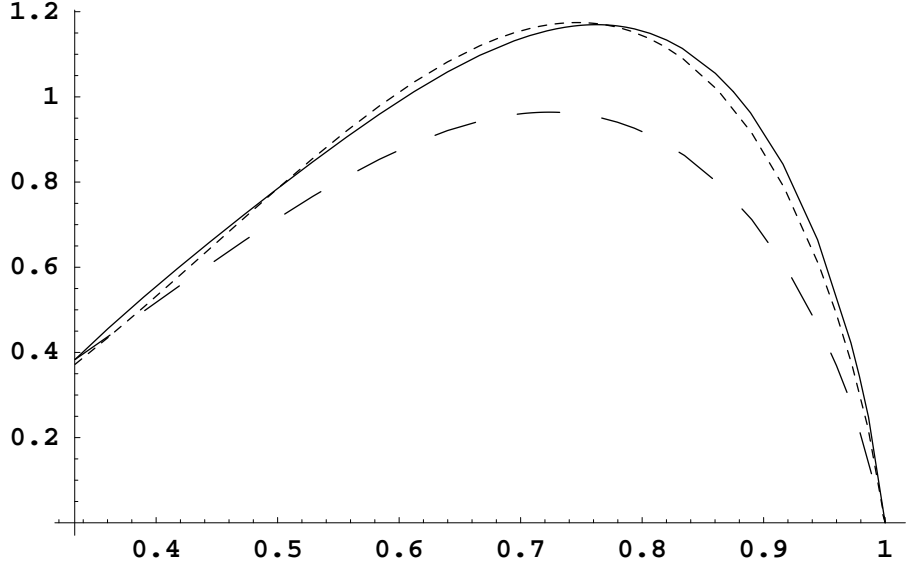}}}
\caption{The entropy difference $4[S(1) - S(0)]$ for the
three channels
\nl $\Phi[\mu, \mu, \mu] \otimes \Phi[\mu, \mu, \mu]$ (dots),
$\Phi[\mu, \mu, \mu] \otimes \Phi[\mu, 2\mu -1, \mu]$ (dashes),
and
\nl $\Phi[\mu, 2\mu -1, \mu] \otimes \Phi[\mu, 2\mu -1, \mu]$ (full)
in the range $\thrd \leq \mu \leq 1$.
\label{fig:entdiff2}}
\end{figure}

\pagebreak

\begin{figure}
\centering{\resizebox{!}{3.5in}{\includegraphics{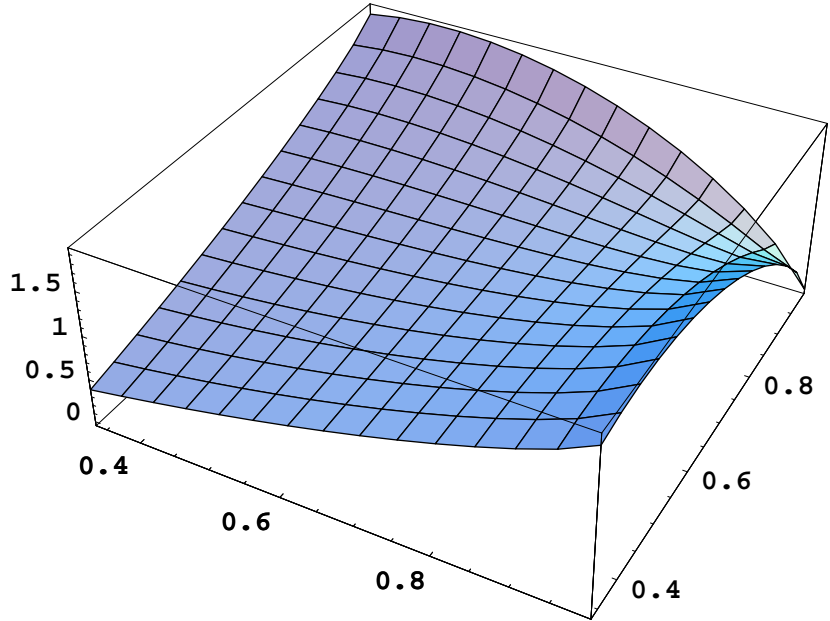}}}
\caption{The entropy difference $4[S(1) - S(0)]$ for the channel
\nl $\Phi[\mu, 2\mu -1, \mu \otimes \Omega[\nu, 2\nu -1, \nu]$
in the range $\thrd \leq \mu, \nu \leq 1$.
\label{fig:entdiff3}}
\end{figure}

\pagebreak

\begin{figure}
\centering{\resizebox{!}{3.5in}{\includegraphics{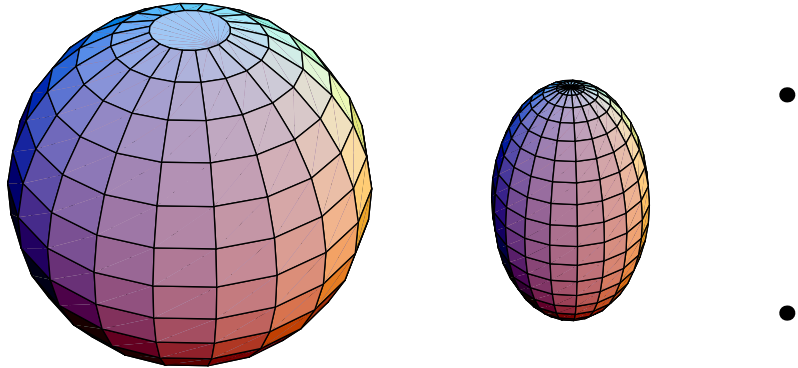}}}
\caption{The Bloch sphere and
its image (the ellipsoid) under a unital map with one large
and two small singular values. The endpoints of the ellipsoid
are shown separately -- these are the unique states of minimal entropy.
\label{fig:ellip1}}
\end{figure}

\pagebreak

\begin{figure}
\centering{\resizebox{!}{3.0in}{\includegraphics{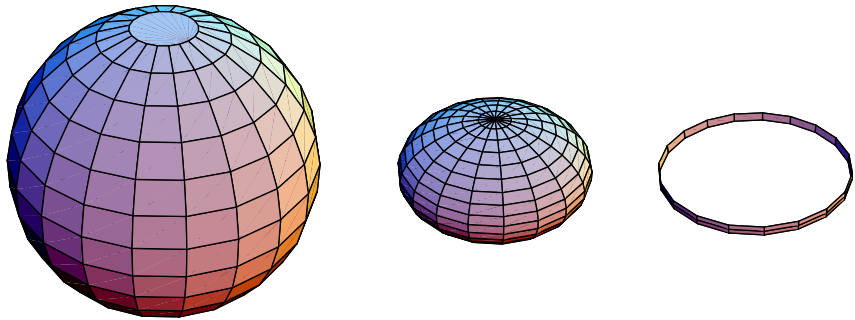}}}
\caption{The Bloch sphere
and its image under a unital
map with two large and one small singular values. The
``waistband'' of the ellipsoid is shown separately;
this entire circle consists of minimal entropy states.
\label{fig:ellip2}}
\end{figure}

\pagebreak

\begin{figure}
\centering{\resizebox{!}{4.0in}{\includegraphics{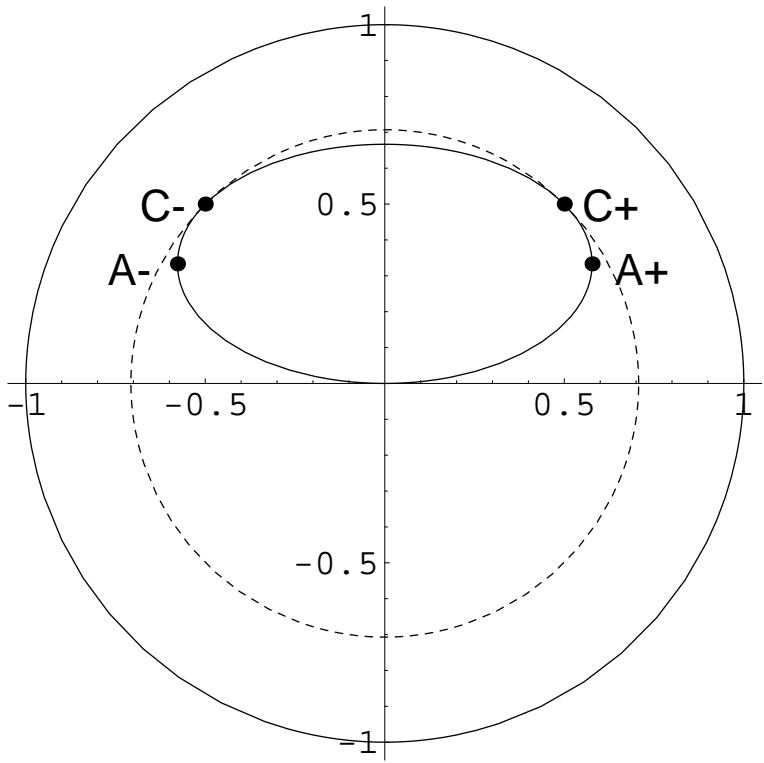}}}
\caption{The Bloch sphere (the circle of
radius 1) and its image (the translated ellipse) under the
Fuchs map, together with the circle of radius
$1/\sqrt{2}$. The endpoints of the ellipse are marked $A
\pm$, and the points of minimal entropy are marked $C \pm$.
\label{fig:Fuchs}}
\end{figure}

\end{document}